\documentclass[twocolumn]{aastex631}

\usepackage{newtxtext,newtxmath}
\usepackage{amsmath}
\usepackage{latexsym}
\usepackage{amssymb}	
\usepackage{gensymb}
\usepackage{flafter}
\usepackage{appendix}
\usepackage{natbib}
\graphicspath{{./}{figures/}}

\usepackage{subfigure}
\usepackage[figuresright]{rotating}
\usepackage{makecell}
\usepackage{booktabs}
\usepackage{multirow}
\usepackage{threeparttable}
\usepackage{color}

\begin{document}

\title{X-Ray spectral and  temporal properties of LMXB 4U 1608-52-- observed with {\it AstroSat} and {\it NICER}}
\shorttitle{X-Ray spectral and temporal analysis of 4U 1608--52}
\shortauthors{S. Bhattacherjee et al.}

\author{Sree Bhattacherjee}
\affiliation{Department of Applied Sciences, Tezpur University, Napaam, Assam 784028, India}
\thanks{E-mail: biplobs@tezu.ernet.in}
\author{Ankur Nath}
\affiliation{Department of Physics, Lovely Professional University, G. T. Road, Phagwara, Punjab 144411, India}
\author{Biplob Sarkar}
\affiliation{Department of Applied Sciences, Tezpur University, Napaam, Assam 784028, India}
\author{Aru Beri}
\affiliation{DST-INSPIRE Faculty, Indian Institute of Science Education and Research (IISER) Mohali, Punjab 140306, India}
\author{Suchismito Chattopadhyay}
\affiliation{Department of Physics, Govt. Girls’ General Degree College, 7 Mayurbhanj Road, Kolkata 700023, India}
\author{Yashpal Bhulla}
\affiliation{Pacific Academy of Higher Education and Research University, Udaipur 313003, India}
\author{Ranjeev Misra}
\affiliation{Inter-University Centre for Astronomy and Astrophysics(IUCAA), Pune, Maharashtra 411007, India}

\begin{abstract}

We report results from a detailed study of the neutron star X-ray binary, 4U 1608-52 using observations with {\it AstroSat} (LAXPC/SXT) and {\it NICER} during its 2016 and 2020 outbursts. The 0.7--20.0 keV spectra could be well described with the disk blackbody and thermal Comptonization model. The best-fitting inner disk temperature is $\sim$ 1 keV and radius { $\sim$ 22.17$^{+2.57}_{-2.38}$--27.19$^{+2.03}_{-1.85}$} km and no significant evolution was observed in the disk radius after performing flux and time-resolved spectroscopy. We used a multi-Lorentzian approach to fit the power density spectra and obtained broad-band noise variability. We estimated the energy-dependent fractional root mean square and time-lag of the broad-band noise, and these variations are quantitatively modelled as being due to the coherent variation of the disk emission and the coronal heating rate. Thus, the rapid temporal modeling is consistent with the longer term spectral evolution  where the inner disk radius does not vary, and instead the variations can be attributed to accretion rate variations which changes the inner disk temperature and the coronal heating rate.
	
\end{abstract}


\section{Introduction} \label{sec:intro}

X-ray binaries are systems comprising of either a black hole (BH) or a neutron star (NS) that accretes matter from a regular star.  If the mass $M$ of the companion star is $ \leqslant 1.0 M_\odot$, the system is termed as low-mass XRBs or LMXBs \citep[refer e.g.][]{2006Lewin}. 4U 1608-52, is an LMXB, harboring a NS \citep{1976Tananbaum}. NS-LMXBs are further classified into atoll and Z sources \citep{1989Hasinger} based on the pattern followed by the hardness-intensity diagram (HID), a plot of the high-energy X-ray photon ratio (hard color) vs. the total intensity \citep{1997Wijnands}. 

4U 1608--52 is found to be an established atoll source \citep{1989Hasinger, 2003Straaten}. The atoll sources { are detected in} two primary states: the hard or lower luminosity state, named as the island state (IS), and the soft or higher luminosity state named as the banana state (BS) \citep{1989vander, 1989Hasinger, 2000vander, 2001Barret, 2006vander, 2014Church}. { Atoll} type sources trace a `C'-like pattern \citep{2007Done} in their HID and color-color diagram (CCD) (a plot of high vs. low energy X-ray photon count ratio) \citep{2001Disalvo}, and continues trailing the path back and forth { overtime}. The underlying reason to form such a track in the HID is indistinct; some { reports} claim that it is due to the  mass accretion rate ($\dot{M}$) variation, which usually increases from IS to BS \citep{1989Hasinger, 1990vander, 1994vander}; while { some recent studies propose that it is not solely a result of $\dot{M}$, but rather of instabilities induced} by the accretion flow at the inner disk radius or the boundary layer of the NS surface \citep{2002Homan, 2002Kuulkers, 2010Homan, 2007Lin, 2009Lin}. The atoll sources { are known to} exhibit striking spectral variations over timescales { spanning days to weeks} \citep{1990vander, 1997Prins, 2002Marek, 2007Lin}. The photon hardness mapping helps to understand the spectral changes; the correlations observed in such mapping can be better studied and parameterized by spectral modelling of the source.

The source 4U 1608-52 is one of the X-ray bursters in the Norma constellation \citep{1976Tananbaum, 1976Belian}. It is estimated to be situated { at a distance} within the range $\simeq$ 2.9--4.5 kpc \citep{2008Galloway, 2014Poutanen}. The spin frequency associated with the central object is 619 Hz \citep{2003Hartman, 2008Galloway}, making it one of the fastest rotating NSs. \citet{2010Guver} estimated the mass and radius of the source to be 1.74 ± 0.14 \(M_\odot\) and 9.3 ± 1.0 km, respectively. 4U 1608-52 is reported to be an X-ray burster \citep{1980Murakami, 1993Yoshida, 2019Jaisawal, 2021Guver, 2022Chen}, similar to other NS-LMXBs such as 4U 1636--53 \citep{1984Turner, 1987Lewin, 2008Galloway, 2019Beri, 2021Roy, 2022Roy}, 4U 1728--34 \citep{2003Shaposhnikov}, Aql X-1 \citep{1998Campana}.

The accretion disk and the corona are considered to be the source of X-ray emission for NS-LMXBs \citep{1989Hanawa,2001Popham,2003Gilfanov}. One needs to investigate the X-ray spectrum of the source as a method to comprehend the accretion process. Two of the widely accepted model types are (1). the Eastern Model \citep{1984Mitsuda, 1986Makishima, 1989Mitsuda}, assuming the softer component is emitted from the disk, and (2). the Western Model \citep{1988White}, where the softer component is characterized as a surface emission. The Comptonized harder component is assumed to result from corona or boundary emission/inner disk in both models \citep{2006Paizis,2006Falanga}. Studying a broad range of RXTE data of { 4U 1608-52}, which was fitted with a blackbody and a Comptonization model, \citet{2002Marek} suggested that the mass accretion rate is the primary cause of the spectral evolution, which is in agreement with the earlier prediction about the change in $\dot{\rm M}$ which increases from IS to BS \citep{1990vander, 1990Hasinger, 1995vander, 1997Prins}. Nevertheless, it is still unclear why exactly such a transition happens. 
\citet{2017Padilla} used the three-component hybrid model in 2010 outburst data of 4U 1608-52 from {\it Suzaku} observation. The thermally Comptonized model was used to account for the hard emission and the blackbody model (single and multi-color) to account for the soft emission from the NS surface and disk. They found the { model} combination to successfully fit the data from soft to hard states. 
\citet{2019Chen} used a thermal Comptonized power-law added to a blackbody to fit the 2018 outburst data. Most recently, \citet{2022Chen} provided insights about how the XRB accretion environment might be related to the thermonuclear burst emission, aiming the study of X-ray spectrum of 4U 1608--52 in PRE burst and persistent emission during its 2020 outbursts.

Determining the power density spectrum (PDS) of the source's light curve in the accessible energy band \citep{1989van} allows searching for variabilities, such as a quasi-periodic oscillation (QPO) or a broad-peaked noise component \citep{1989vanderklis, 1994vander,1994Van, 1997Vaughan}. We can determine the radiative origin of the source's observed variability by calculating the energy-dependent phase-lags and fractional root mean square (rms). 

To explain the observed energy-dependent fractional rms and time-lag for kHz QPOs, a Comptonization model has been developed which considers the presence of a feedback loop, where a fraction of varying high energy photons impinge back on the soft photon source, leading to a varying seed photon input \citep{2001Lee, 2013Avellar, 2014Kumar}. This formalism has been extended to lower frequency QPO \citep{2022Bellavita} and provides estimates of the size of the Comptonizing region. Another interpretation of the time-lags are due to fluctuations propagating inward in a disk \citep{1997Lyubarskii}. If higher-energy photons are produced, preferentially in the inner regions, such propagation will lead to the observed hard time lags \citep{2000Misra, 2001Kotov}. An extension of the model is that the fluctuations propagate from the disk to the corona after a time delay, leading to a variation of the disk emission, followed by a change in the heating rate of the corona \citep{2019Maqbool, 2019Jithesh, 2020Mudambi}. While these works were limited to predicting the variability of the Comptonized component, \citet{2020Garg} have further extended the formalism to take into account the cases when the disk emission is also observed in the spectra.

The Large Area X-ray Proportional Counter (LAXPC) and the Soft X-ray Telescope (SXT) onboard {\it AstroSat}, observed 4U 1608-52 twice during the decay phase of its outburst. In this work, we report the first analysis of these observations along with a simultaneous  Neutron Star Interior Composition Explorer ({\it NICER}) observation. The combination of LAXPC and {\it NICER} data provide a unique opportunity to study the temporal properties of the system in a wide energy band. LAXPC and {\it NICER} have good time resolution (refer Table \ref{tab:tab1}), together it give us a broad energy range (0.2–80.0 keV) to analyze the timing properties among all the instruments combinations. The goal of this work is to use these observations to perform a detailed spectral analysis along with temporal studies of the source, which is a convenient way to understand the accretion phenomenon of the XRBs.

Our paper is organized as follows: the observation and data reduction process that we used to extract the scientific data is covered in  Section~\ref{sec:Obs and data}, Section~\ref{sec:temporal} represents the temporal analysis { where we deal with the lightcurves and HIDs}, Section~\ref{sec:SA} and Section~\ref{sec:TA} deal with the broad-band spectral and timing analysis respectively { and finally}, we discuss and summarize our result in Section~\ref{sec:disc}.

\section{Observation \& Data Reduction} \label{sec:Obs and data}

There are two archival {\it AstroSat} data sets available for 4U 1608-52, which were taken on August 28, 2016 (hereafter set-I) and July 14, 2020 (hereafter set-II), respectively. The second observation has two simultaneous {\it NICER} observations that we will utilise for broad-band analysis. The catalogue of the observations used for the study is given in in Table~~\ref{tab:tab1}.

\subsection{{\textbf{\it  AstroSat }}}
{\it AstroSat} is the first multi wavelength (optical to hard X-ray) space observatory in India \citep{2006Agrawal, 2014Singh}. Below are brief descriptions of the two {\it AstroSat} payload instruments (LAXPC and SXT) that were used for our study.
\begin{figure}
\centering
	\includegraphics[width=1.1\linewidth, height=0.40\textwidth]{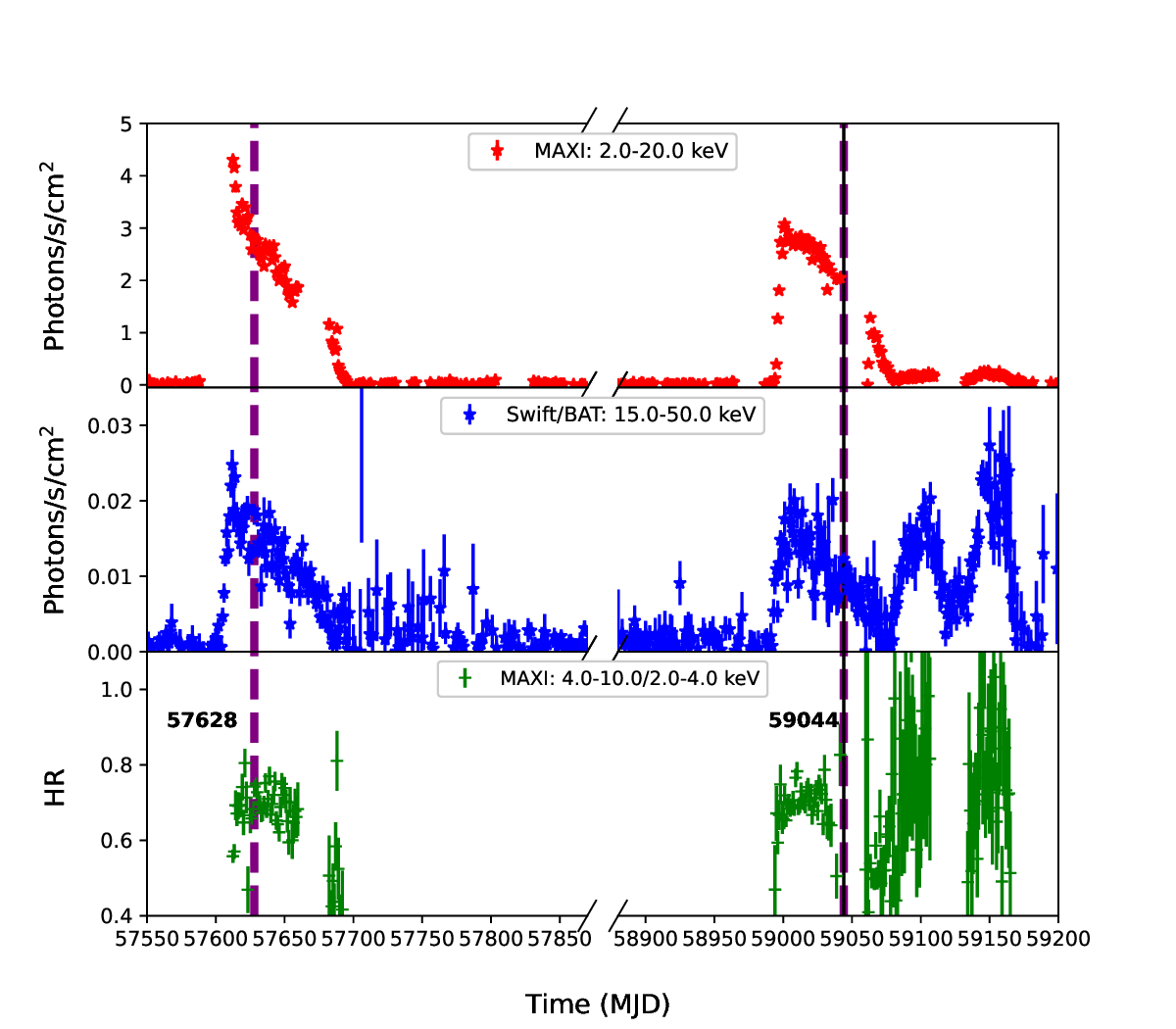}
    \caption{The long-term monitoring of X-ray outbursts from 4U 1608-52 by the instruments MAXI - 2-20 keV (top panel) and  Swift/BAT - 15-50 keV (middle panel). The bottom panel shows the MAXI light curve’s hardness ratio (HR) for an energy band ratio of 4.0-10.0 keV/2.0-4.0 keV. The dashed vertical lines mark the times when {\it AstroSat} observations were taken: 28 August 2016 (MJD 57628) and 14 July 2020 (MJD 59044). The {\it NICER} observation, shown using the thin, solid vertical line, was also taken on 14 July 2020 (MJD 59044).}
    \label{fig:maxi}
\end{figure}

\begin{figure*}
\centering
	\includegraphics[scale=0.33, angle=-90]{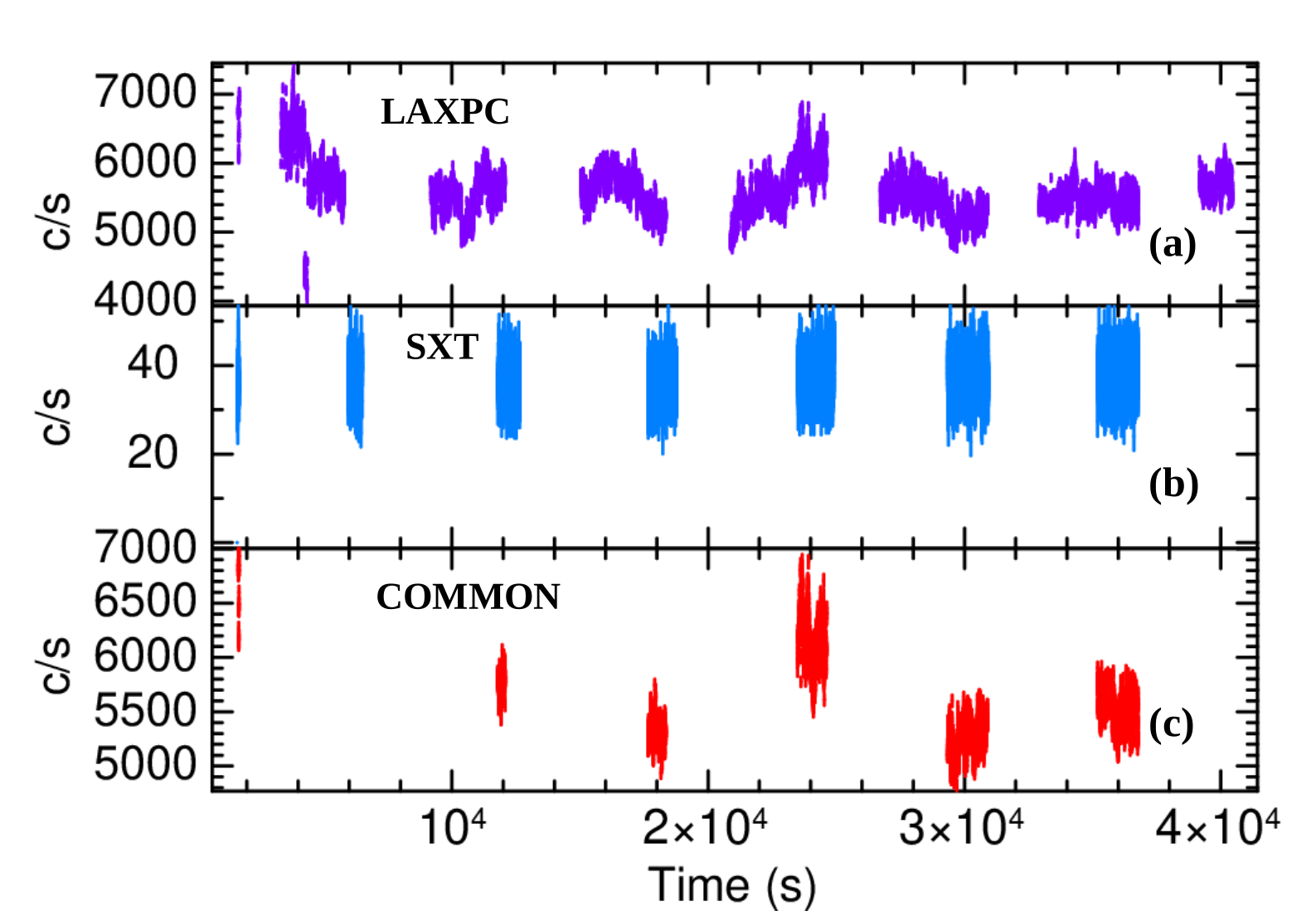}
	\includegraphics[scale=0.33, angle=-90]{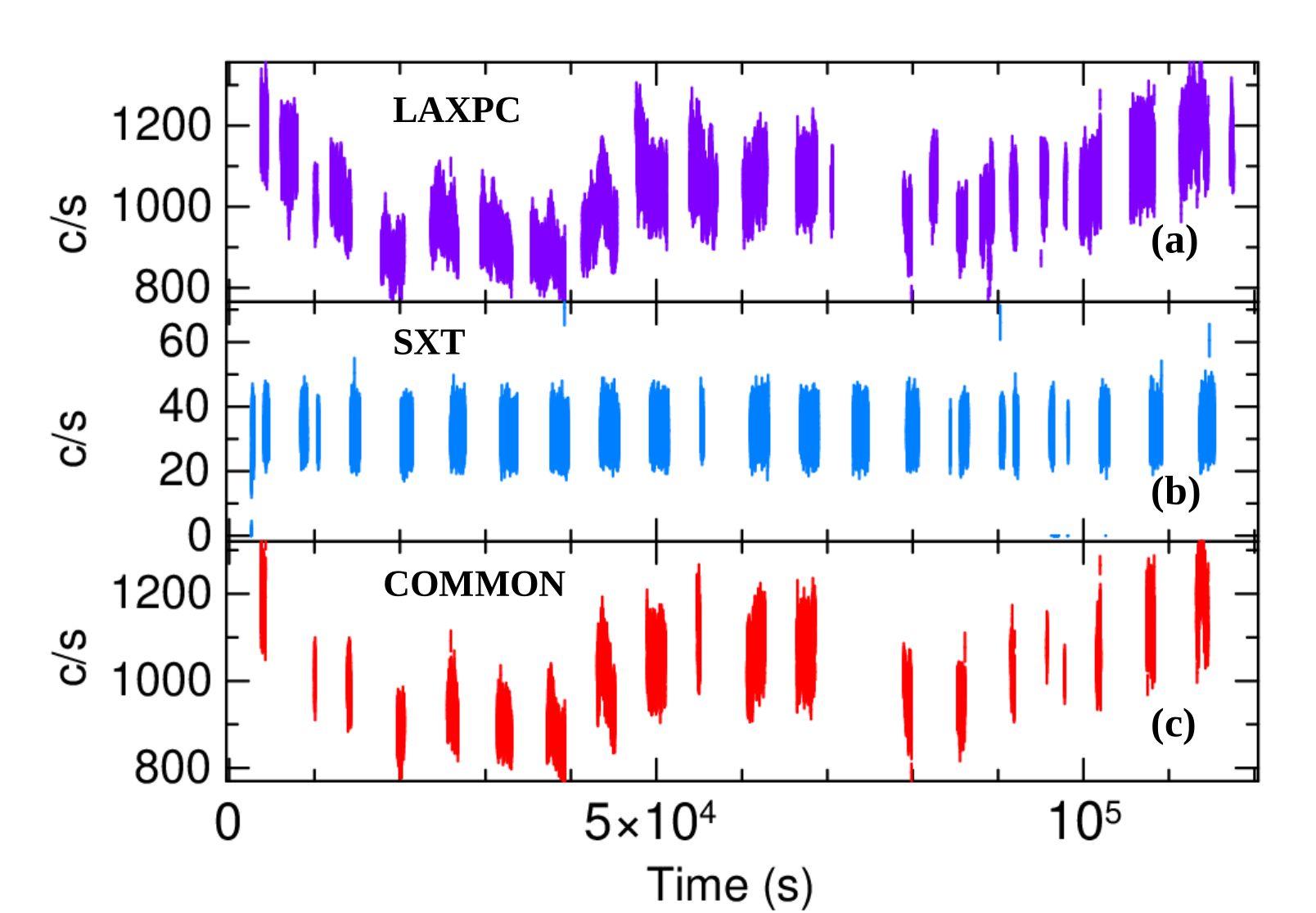}	
    \caption{The background-subtracted light curves for the data set-I (left) and set-II (right) from the instruments LAXPC and SXT of {\it AstroSat}  are shown. For both plots, LAXPC light curves are at the top panel, SXT light curves are at the middle panel, and the light curves obtained for the duration of simultaneous observation by LAXPC and SXT are at the bottom panel. We have used 2.3775 s binsize for plotting these light curves. The LAXPC and SXT light curves are generated for 3-20 keV and 0.3-8 keV, respectively.}
        \label{fig:lc_astrosat}
\end{figure*}

\begin{table*}
\caption{\label{tab:tab1} { Log table for the observations of} 4U 1608$-$52. }
\scalebox{0.80}{
\begin{tabular}{cccccccc}
\hline\hline
 Instrument & Set & Obs ID & Date (YYYY-MM-DD) & Exposure (ks) & Radius (arc min) & Energy range (keV) & Time resolution\\ \hline
\hline
 {\it AstroSat}/LAXPC  & Set-I    & G05\textunderscore 140T01\textunderscore9000000628	        & 2016-08-28  & $\sim$ 40  & --& 3.0-80.0 & 10 $\mu s$ \\
 {\it AstroSat}/SXT    & Set-I    & G05\textunderscore 140T01\textunderscore9000000628 	& 2016-08-28  & $\sim$ 35  & 6 \& 12 & 0.3-8.0& 2.3775 s\\

{\it AstroSat}/LAXPC	& Set-II    & T03\textunderscore 214T01\textunderscore9000003758 	& 2020-07-14  & $\sim$ 51  & -- &3.0-80.0 & 10 $\mu s$\\
{\it AstroSat}/SXT    & Set-II    & T03\textunderscore 214T01\textunderscore9000003758 	& 2020-07-14  & $\sim$ 30  & 5 \& 15 & 0.3-8.0 & 2.3775 s\\
\hline
{\it NICER}/XTI        &    -     &                       3657024001                   	& 2020-07-14  & $\sim$ 0.58 & -- & 0.2-12.0 & { 100} ns\\
{\it NICER}/XTI        &    -     &                       3657024101                  	        & 2020-07-14  & $\sim$ 3.70  & -- & 0.2-12.0 & { 100} ns\\

\hline
\end{tabular}}
\end{table*}

LAXPC has three identical proportional counters units (PCUs) 10, 20, and 30, with an operational energy range of 3--80 keV. LAXPC uses the event mode data and has a fine time resolution of 10 $\mu$s \citep{2016Yadav, 2017Agrawal, 2017Antia}. In our work for set-I, we have used all three counters for timing analysis, and only LAXPC 10 and 20 for spectral analysis of the X-ray data, but for set-II, we restricted ourselves to LAXPC 20 data only. Because LAXPC 10 has been operating at low gain since March 28, 2018, and LAXPC 30 was switched off on March 8, 2018, due to abnormal gain changes. We modelled the spectrum in the 4.0--20.0 keV energy band.

The data reduction and analysis are carried out by the standard tools provided by the latest version of LAXPCsoftware\footnote{\url{ http://astrosat-ssc.iucaa.in/laxpcData}} (version: Oct. 13, 2022). The software was used to extract the scientific Level-2 data using the good time interval (GTI) data uninfluenced by Earth occultation and the South Atlantic Anomaly (SAA) \citep{2017Agrawal}. The standard procedure as per the ASSC website\footnote{\url {http://astrosat-ssc.iucaa.in/uploads/threadsPageNew_SXT.html}} is used to extract the scientific results, using the corresponding background and response files for the counters. The {\tt as1bary tool} is utilised for barycenter correction of the data (set-II). The uppermost layer (L1) data has been considered in order to reduce the background (for more details, refer to \citet{2021Antia, 2019Beri, 2022Nath}). 

\begin{figure}
	\centering
	\includegraphics[scale=0.5]{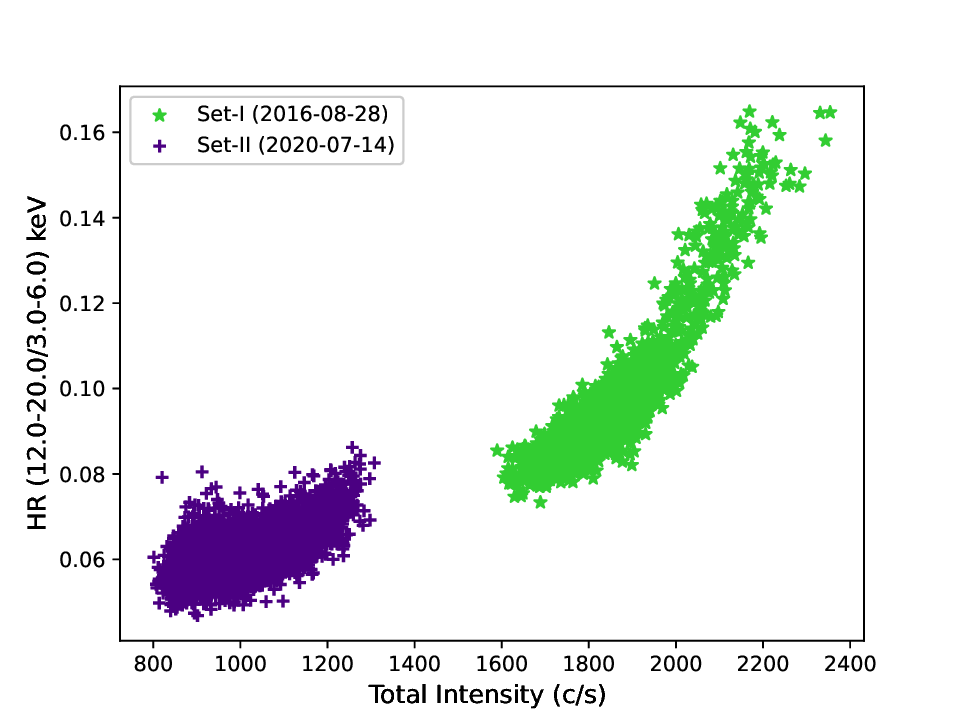}
    \caption{The Hardness Intensity Diagram (HID) of 4U 1608-52, as generated from {\it AstroSat}/LAXPC20 unit data, are shown in the figure. The binning size of the data point counts to 16 s, and the total intensity range is measured for the energy band 3-20 keV in units of c/s.}
    \label{fig:HID_astrosat}
\end{figure}

\begin{figure*}
\includegraphics[scale=0.59]{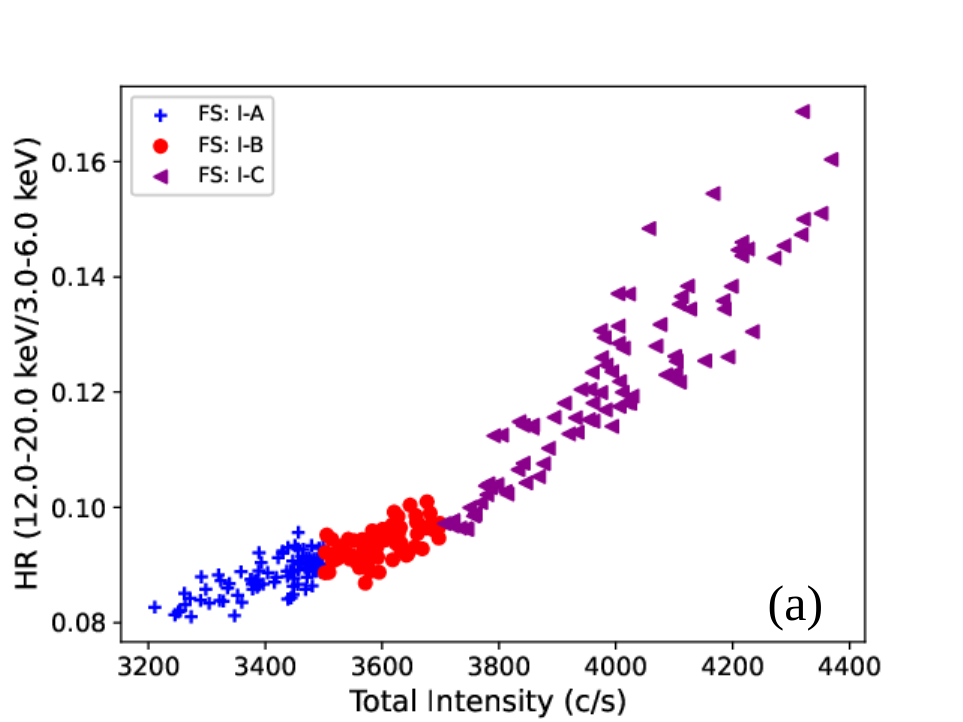}
\includegraphics[scale=0.59]{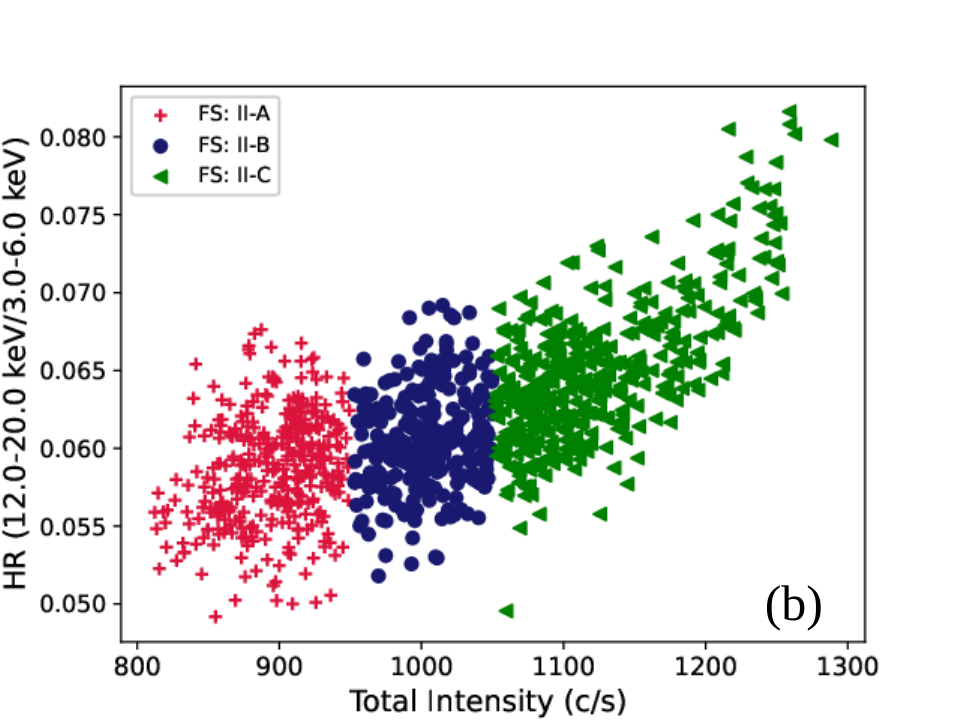}
\caption{\label{fig:ccdhid} The HID of 4U 1608-52 obtained with the strict simultaneous LAXPC/SXT data for both the {\it AstroSat} data sets; (a) the three flux levels I-A (3200-3500 c/s), I-B (3500-3700 c/s), and I-C (3700-4400 c/s) of set-I, (b) the three flux levels II-A (800-950 c/s), II-B (950-1050 c/s), and II-C (1050-1300 c/s) of set-II.}
\end{figure*}


SXT is one of the X-ray instruments onboard {\it AstroSat} that is sensitive to a soft energy band of 0.3--8.0 keV \citep{2016Singh, 2017Singh} with a large time resolution of $\sim$2.3775s, used especially for X-ray imaging and spectroscopy. The level-1 SXT data for the source has been processed through the sxtpipeline {\tt (AS1SXTLevel2-1.4b)} to generate the cleaned level-2 event files, every SXT observation is made in Photon Counting (PC) mode. Using the SXT Event Merger Tool (Julia), the level-2 event files from different orbits were merged, which was used to extract the source lightcurve, spectra and image using the {\sc XSELECT V2.5b} of {\sc HEASOFT} (v. 6.31.1). A gain fit correction was used to modify the response while modelling, keeping the slope fixed at one and leaving the offset to vary. An offset correction of 0.02-0.09 is needed in quite a few SXT observations \citep{2021Beri, 2023Beri}. The spectrum has been binned using the optimal binning approach of ftool, {\tt ftgrouppha} \citep{2016Kaastra}. The source count rate was much higher than the pileup threshold value (i.e., >40 c/s) in photon counting mode\footnote{\url{https://www.tifr.res.in/~astrosat_sxt/dataanalysis.html}}, showing possibility of a pileup in the energy spectrum. Until the source count rate fell below the threshold rate for the pileup, we eliminated the source counts from the centre region of the point-spread function (PSF) in order to reduce the pileup effect. We found that for an annulus region with an inner radius and outer radius mentioned in Table~\ref{tab:tab1}, the source count rate becomes lower than the aforesaid threshold value. For the present source region, a vignetting corrected ancillary response files (ARF) has been generated for the respective segments using the {\sc sxtARFModule}. For background estimation and response of the instrument, the standard spectrum {\tt SkyBkg\_comb\_EL3p5\_Cl\_Rd16p0\_v01.pha}, and response matrix file (RMF) {\tt sxt\_pc\_mat\_g0to12.rmf} has been used as provided by the SXT team\footnote{\url{http://astrosat-ssc.iucaa.in/sxtData}}.

\subsection{{\textbf{\it NICER }}}

{\it NICER}, a payload instrument of the International Space Station (ISS), was launched in 2017. The X-ray timing instrument (XTI) is present at the heart of the {\it NICER}, consisting of 56 X-ray optics and silicon drift detector (SSD) pairs \citep{2012Gendreau}, of which 52 are in functioning condition\footnote{\url{https://heasarc.gsfc.nasa.gov/docs/nicer/data_analysis/workshops/NICER2021workshopwelcome.pdf}}, offer a high time resolution of { 100} ns\footnote{\url{https://heasarc.gsfc.nasa.gov/docs/nicer/mission_guide/}} \citep{2022Remillard}. {\it NICER} works in the energy range of 0.2-12 keV \citep{2014Arzoumanian}. 

The {\it NICER} data are reduced using the standard {\tt nicerl2} pipeline\footnote{\url{https://heasarc.gsfc.nasa.gov/lheasoft/ftools/headas/nicerl2.html}}, excluding the noisy detectors 14 and 34\footnote{\url{https://heasarc.gsfc.nasa.gov/docs/nicer/data_analysis/nicer_analysis_tips.html}}. We used the {\it NICER} CALDB version of 20221001\footnote{\url{https://heasarc.gsfc.nasa.gov/docs/heasarc/caldb/nicer/}}. For the two data sets, a merged clean event file and GTI has been created, {\tt nibackgen3c50} has been used with gain epoch 2020 for the background estimation \citep{2022Remillard}. To generate the {\it NICER} spectra, the optimal binning algorithm of ftool `{\tt ftgrouppha}' is used \citep{2016Kaastra}. The ftools {\tt nicerrmf} and {\tt nicerarf} are used to create the RMF and ARFs.  The {\sc XSELECT V2.5b} has been further used to extract the science products of {\it NICER}, such as the lightcurve and spectra. 


\section{Lightcurves and Hardness Ratios}
\label{sec:temporal}

The {\it AstroSat} and {\it NICER} observations that have been used in this study are shown in Figure~\ref{fig:maxi}, which is a long-term MAXI (2–20 keV) and Swift/BAT (15–50 keV) lightcurves. The observations were made during the decay phase of the 2016 and the 2020 outburst. The lower panel of Figure~\ref{fig:maxi} represent the evolution of hardness ratio of the source, plotted using MAXI long-term lightcurve, filtering the data using S/N ratio of greater than 3. Figure~\ref{fig:lc_astrosat} displays the lightcurve for the LAXPC and SXT instruments in the panel (a) and (b), respectively, and the panel (c) represents the lightcurve for that period of time when both LAXPC+SXT were simultaneously active in each observation. Figure~\ref{fig:HID_astrosat} is the HID of set-I and II, which shows that the source was in the BS during observations. The classification of atoll sources suggests that set-I was in UBS, while set-II was in the LBS during their respective observations \citep{1989Hasinger}.
From Figure~\ref{fig:HID_astrosat} we can see the hardness ratio (HR) (X-ray photon count ratio in 12--20 keV over the 3--6 keV energy range) is varying significantly with the intensity in the BS. This positive correlation in HR led us to investigate the spectral change patterns, by applying flux-resolved spectroscopy--a modelling technique where each spectrum file is generated for small flux ranges (Figure~\ref{fig:ccdhid}) to perform the detailed spectral analysis. We divided the entire flux into three dominant flux { levels (FL)} for each set. First, the GTI files are created by clipping off the fluxes using the LAXPC routine, {\it laxpc\_fluxresl\_gti}. For data set-I, the flux ranges are: (1). 3200-3500 c/s (first {  FL} of set I or I-A) (2). 3500-3700 c/s (second {  FL} of set I or I-B) (3). 3700-4400 c/s (third {  FL} of set I or I-C) [refer Figure~\ref{fig:ccdhid}(a)]. These GTI files are used to generate spectra and the corresponding background spectra for each FS. Similarly, the {  FL} for data set-II are as follows: (1). 800-950 c/s (first {  FL} of set II or II-A) (2). 950-1050 c/s (second  {  FL} of set II or II-B) (3). 1050-1300 c/s (third {  FL} of set II or II-C) [refer Figure~\ref{fig:ccdhid}(b)].

\begin{figure*}
\centering
\includegraphics[width=0.5\linewidth, height=0.7\textwidth, angle=-90]{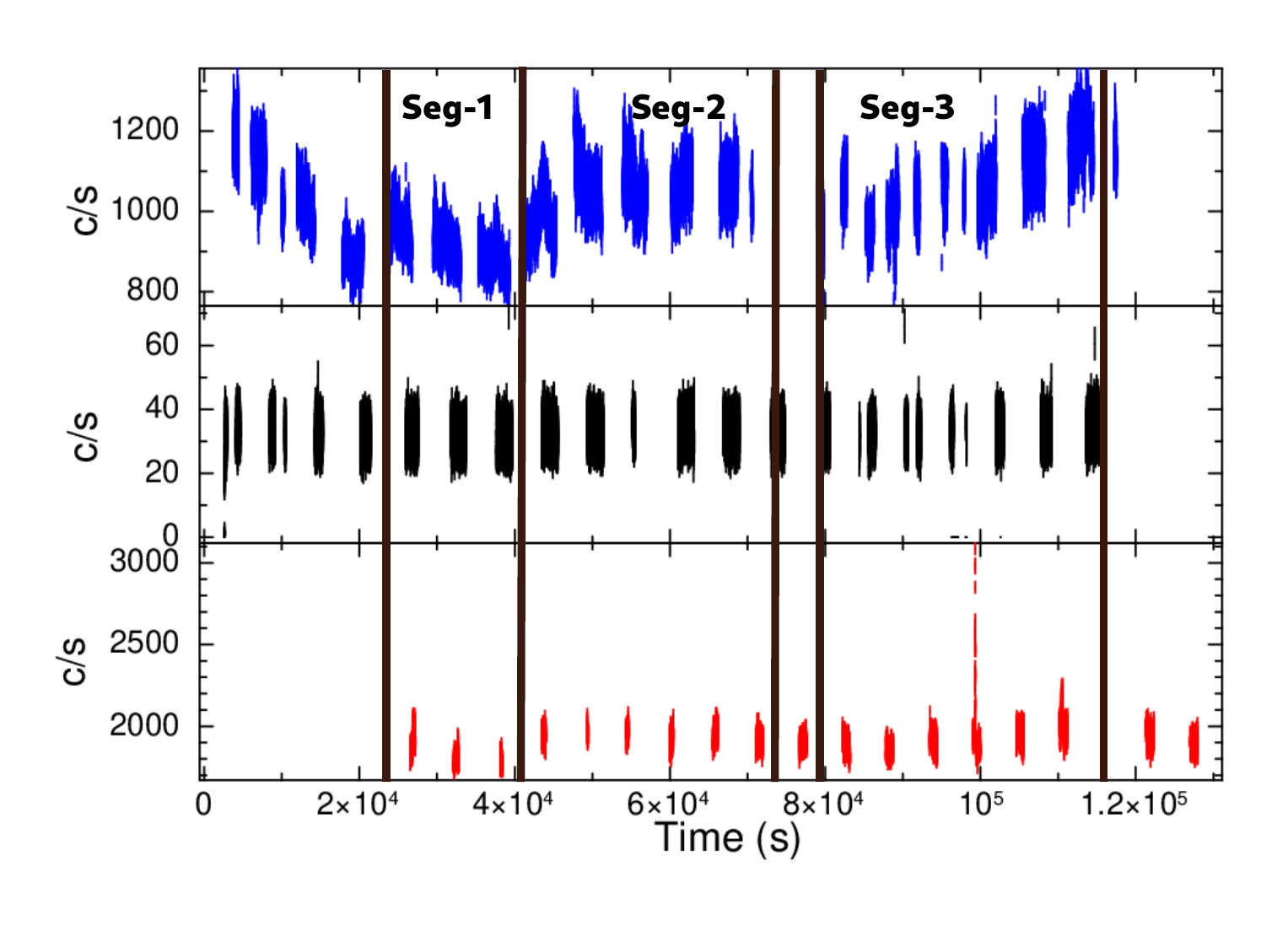}
    \caption{  The barycentric corrected lightcurve of data set-II, using LAXPC 20 {(top)}, SXT {(middle)}, and its simultaneous {\it NICER} data {(bottom)}, in the energy range 3-20 keV, 0.3-8 keV, and 0.5-12 keV, respectively. The binning time used is 2.3775 s. { A thermonuclear burst has been detected at} 10$^5$ s after the start of the {\it NICER} observation.}
    \label{fig:lc_nicer}
\end{figure*}

Figure~\ref{fig:lc_nicer} represents the barycentric corrected lightcurve for {\it AstroSat} (set-II) and {\it NICER} observation, from top to bottom panels, respectively. In the {\it NICER} lightcurve a thermonuclear burst has been detected at $\sim$10$^5$ s after the start of the observation. On investigating the simultaneous segment of the burst with LAXPC and SXT, it was noticed that during that time, both the {\it AstroSat} instruments were inactive, which is why LAXPC, despite having high timing resolution, it missed the burst. With the simultaneous observation of {\it AstroSat} (set-II) and {\it NICER}, we segmented their GTI into three parts as shown in Figure~\ref{fig:lc_nicer}, considering the LAXPC, SXT, and {\it NICER} data and performed time-resolved spectroscopy to probe the source's behaviour along the intervals. 

\begin{figure}
\centering
 \includegraphics[scale=0.50]{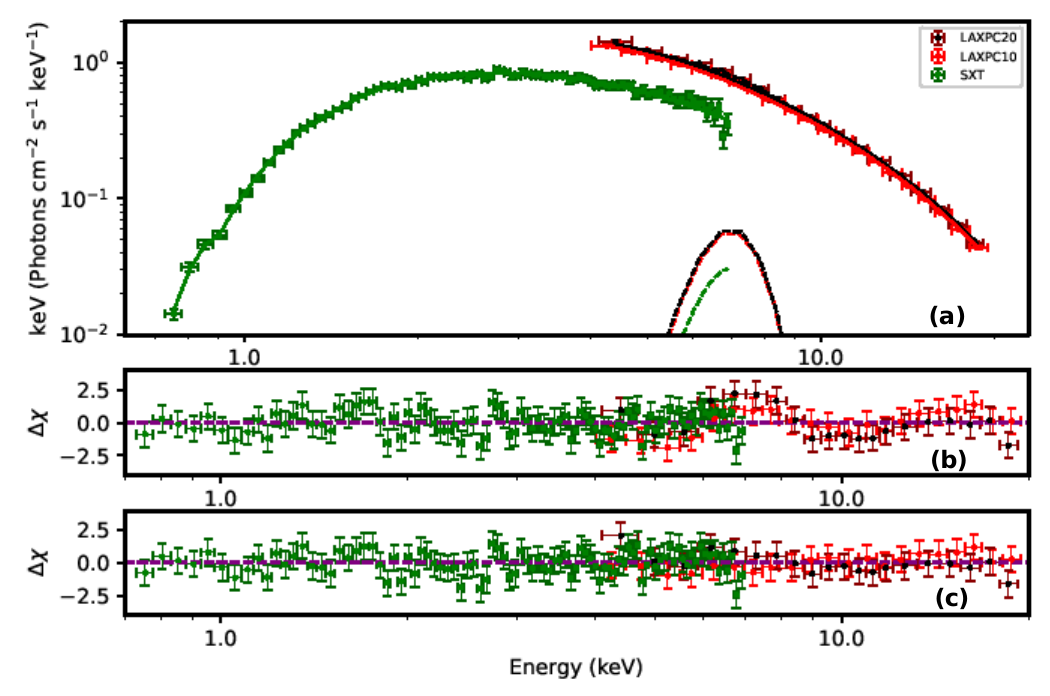}
\caption{Flux-resolved spectroscopy:  The representative spectra for flux level I-A for the X-ray spectrum over the energy range of 0.7–20.0 keV. The panels (a) and (c) show the spectra fitted using const*tbabs*(gauss+thComp*diskbb) model and its corresponding residuals, respectively, whereas the panel (b) shows the residual emission using the continuum model only.}
\label{fig:6spec}
\end{figure}


\section{Broad-band spectral analysis}
\label{sec:SA}
\subsection{Flux--resolved Spectroscopy}
\label{sub:FRS}

Utilizing the broad-band spectral coverage of LAXPC and SXT onboard {\it AstroSat} over the data sets I and II, we carried out the joint-spectral fitting considering the strict simultaneous data of LAXPC and SXT as shown in the bottom panel of Figure~\ref{fig:lc_astrosat}. The LAXPC spectrum has been generated within the energy 4.0--20.0 keV; since above  $\sim$20.0 keV, the background dominates over the source spectra (see Figure~\ref{fig:apdx1}). Whereas, for SXT, the energy range considered for generation of spectrum is 0.7--7.0 keV. A systematic uncertainty of 2\%\footnote{\url{https://www.tifr.res.in/~astrosat_sxt/dataana_up/readme_sxt_arf_data_analysis.txt}} was introduced in the model while fitting. We use the {\sc XSPEC} tool version 12.13.0c \citep{1996Arnaud}, to carry out the spectral modelling.

\begin{table*}
	\centering
	\setlength{\tabcolsep}{8.0pt}
	\caption{{ Flux-resolved spectroscopy:}  { Best-fit} spectral parameters { of the spectra} for the {\it AstroSat} { observations.}}
	\label{tab:tab2}
	\begin{tabular}{@{}cccccccc@{}} 
		\hline
		Spectral Component& Parameter & {I-A} & {I-B} &{I-C} & {II-A} & {II-B} & {II-C}\\
		
		\hline
{\tt Tbabs}& $N_{\rm H}$ ($10^{22}$~cm$^{-2}$)&0.91$^{+0.02}_{-0.02}$ & 0.93$^{+0.03}_{-0.02}$ &0.92$^{+0.03}_{-0.03}$ & 0.97 $^{+0.03}_{-0.03}$   &0.95 $^{+0.02}_{-0.02}$ &   0.95 $^{+0.02}_{-0.02}$\\
\hline
&&&&&&&\\

{\tt gauss}& Line$_{\rm gauss}(\rm keV) $&  6.9$^{*}$  &6.9$^{*}$   &6.9$^{*}$&6.9$^{*}$& 6.9$^{*}$   &6.9$^{*}$\\
&&&&&&&\\
&$\sigma (\rm keV)$&0.81 $^{+0.32}_{-0.32}$ & 0.90 $^{+0.29}_{-0.32}$ & 0.93  $^{+0.35}_{-0.40}$&0.59$^{+0.40}_{-0.28}$  & 0.78 $^{+0.31}_{-0.31}$ & 0.84  $^{+0.35}_{-0.30}$\\ 
&&&&&&&\\
&N$_{\rm gauss}\times 10^{-2}$ &1.87 $^{+0.67}_{-0.75}$& 2.41 $^{+0.78}_{-0.84}$&  2.43 $^{+1.02}_{-0.92}$&0.70$^{+0.36}_{-0.30}$& 1.10$^{+0.47}_{-0.41}$&  1.12 $^{+0.53}_{-0.45}$\\
\hline
&&&&&&&\\

{\tt thComp}& $\Gamma $ &2.55 $^{+0.14}_{-0.11}$& 2.40  $^{+0.10}_{-0.08}$& 2.25$^{+0.09}_{-0.07}$& 2.86 $^{+0.07}_{-0.06}$& 2.85  $^{+0.13}_{-0.11}$& 2.86 $^{+0.14}_{-0.12}$ \\ 
&&&&&&&\\
& kT$_e$ (keV) & 3.23 $^{+0.23}_{-0.16}$ & 3.08  $^{+0.16}_{-0.12}$ &3.14 $^{+0.15}_{-0.11}$ & 4.71 $^{+0.49}_{-0.37}$ & 4.56  $^{+0.76}_{-0.54}$ & 4.60 $^{+0.83}_{-0.50}$\\ 		
\hline
&&&&&&&\\
{\tt diskbb} &T$_{\rm in}$ (keV)&1.25 $^{+0.08}_{-0.07}$  & 1.19  $^{+0.08}_{-0.07}$& 1.22  $^{+0.09}_{-0.08}$& 0.99 $^{+0.01}_{-0.01}$  & 1.08  $^{+0.05}_{-0.05}$&1.15  $^{+0.05}_{-0.05}$\\
&&&&&&&\\
&N$_{\rm dbb} $& 283 $^{+70}_{-57}$& 345 $^{+87}_{-71}$ & 313 $^{+87}_{-69}$& 426 $^{+66}_{-56}$& 322 $^{+56}_{-48}$ & 278 $^{+45}_{-39}$\\  
  &&&&&&&\\ 
  \hline 
& $\chi^2/\rm dof$   & 82.35/123  & 98.12/123 & 102.83 /122& 93.78/107  & 87.29 /104 & 96.71/107\\
             &&($\sim$ 0.67)&($\sim$ 0.80)&($\sim$ 0.84)&($\sim$ 0.88)&($\sim$ 0.84)&($\sim$ 0.90)\\
             &&&&&&&\\
		\hline
		
&Inner disk radius &7.67 $^{+0.89}_{-0.82}$& 8.46  $^{+1.01}_{-0.92}$& 8.06 $^{+1.06}_{-0.94}$&9.41 $^{+0.70}_{-0.64}$& 8.18  $^{+0.68}_{-0.64}$& 7.60 $^{+0.59}_{-0.55}$ \\
&R$_{\rm in}$  (km) & & &\\
		\hline
&R$_{\rm in (co)}$ (km)&22.17$^{+2.57}_{-2.38}$&24.45$^{+2.91}_{-2.66}$&23.29$^{+3.07}_{-2.71}$& 27.19$^{+2.03}_{-1.85}$& 23.64$^{+1.70}_{-1.85}$&21.96$^{+1.71}_{-1.59}$\\

\hline

 & Total unabsorbed     & 2.14 $^{+0.05}_{-0.05}$  & 2.24 $^{+0.05}_{-0.05}$ & 2.40 $^{+0.05}_{-0.06}$& 1.17 $^{+0.03}_{-0.02}$  & 1.26 $^{+0.03}_{-0.03}$ & 1.38 $^{+0.03}_{-0.03}$ \\
 &flux (10$^{-8}\rm ergs/\rm cm^{2}/s$)& & &\\
 
 \hline
& Disk flux   & 1.32 $^{+0.06}_{-0.06}$  & 1.32 $^{+0.06}_{-0.06}$ & 1.32 $^{+0.06}_{-0.06}$ & 0.74 $^{+0.03}_{-0.03}$  & 0.81 $^{+0.04}_{-0.04}$ & 0.89$^{+0.06}_{-0.04}$\\
  & (10$^{-8}\rm ergs/\rm cm^{2}/s$)& & &\\
  
  \hline

 & Mass accretion rate ($\dot{\rm M}$)    & 2.97 $^{+0.07}_{-0.07}$  & 3.11 $^{+0.07}_{-0.07}$ & 3.33 $^{+0.07}_{-0.08}$& 1.62 $^{+0.03}_{-0.04}$  & 1.75 $^{+0.04}_{-0.04}$ & 1.92 $^{+0.04}_{-0.03}$ \\
   & (10$^{17}\rm gm/\rm s$)& & &\\
  \hline\hline
{$^*$Frozen parameter.}\\
\multicolumn{8}{p{18cm}}{Notes: $N_{\rm H}$ is neutral hydrogen column density; $\sigma$ is Gaussian line width; $\Gamma$ is asymptotic power-law index; kT$_e$, T$_{\rm in}$ represent the electron and inner disk temperature respectively; $\chi^2/\rm dof$ is $\chi^2$ per degrees of freedom, the values below them are $\chi^2_{\rm red}$ value. R$_{\rm in(co)}$ (km) is color-corrected inner disk radius.}
\end{tabular}		
\end{table*}

\begin{figure*}
\centering
\includegraphics[scale=0.49, angle=0]{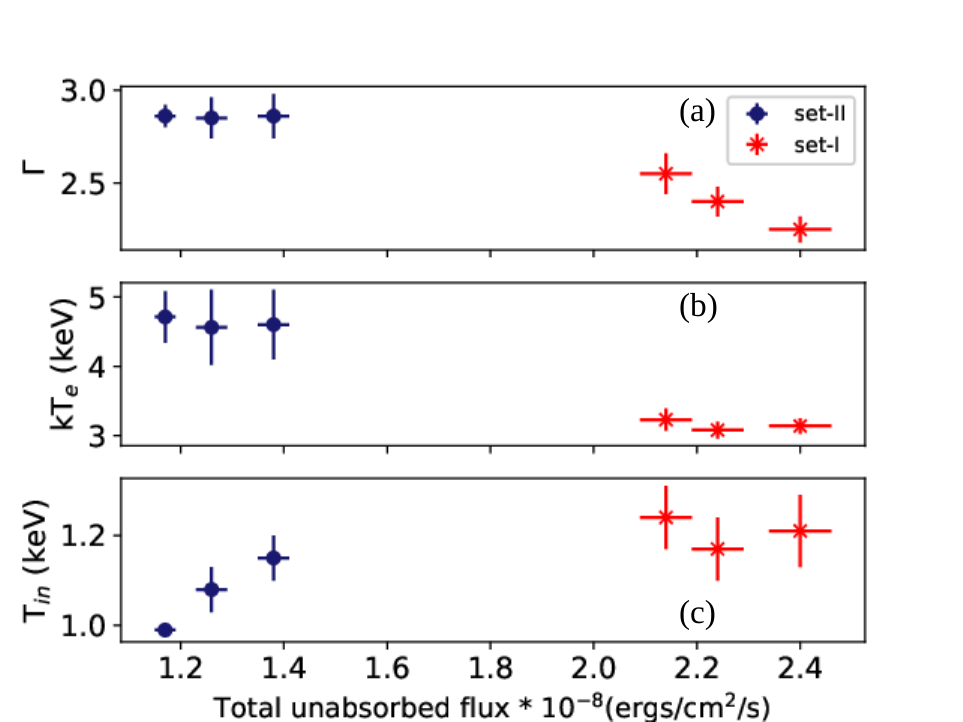}
\includegraphics[scale=0.49, angle=0]{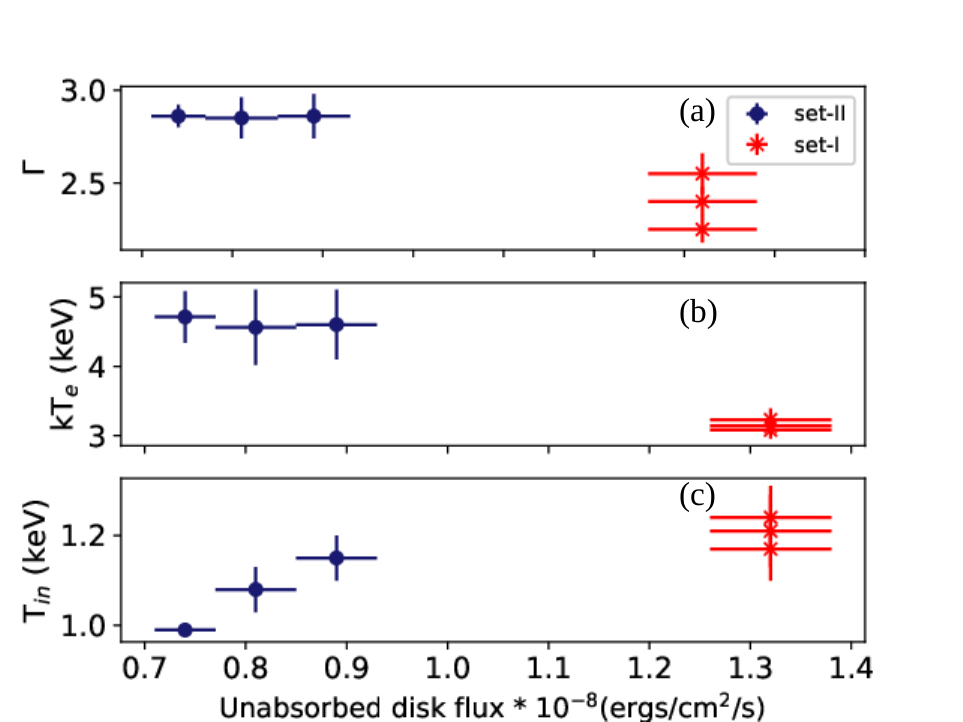}
	
    \caption{\label{fig:flux} Variation of spectral parameters $\Gamma$, kT$_e$ and T$_{\rm in}$  with the total unabsorbed flux (left) and unabsorbed disk flux (right) for the two {\it AstroSat} data sets. } 

\end{figure*}

The joint-spectral fitting has been taken up in all six { FL} along the HID track. We start modelling the broad-band (0.7--20.0 keV) X-ray spectrum with the simple model {\tt const*tbabs*(gauss+thComp*bbodyrad)} considering boundary layer of the compact object to be the source of the seed photons that are being Comptonized. To account for the thermal Comptonization, {\tt thComp} model \citep{2020Zdziarski} was used. The Tuebingen-Boulder ISM absorption model, {\tt tbabs} has been included \citep{2000Wilms} to account for the galactic absorption. Throughout the paper, we assumed the distance of the source to be 4 kpc \citep{2010Guver}. { \citet{2010Guver} estimated the distance of the LMXB source 4U 1608-52 using a Gaussian function in order to fit the distribution, and found the best-fit position at 5.8$^{\rm +2.00}_{\rm -1.80}$ kpc, where the negative uncertainty corresponds to the lower-end cutoff, and the positive uncertainty represents the standard deviation of the Gaussian function. The authors repeated the same analysis using Chandra observation, which results the highest possible distance of the source to be $\sim$ 4 kpc \citep{2016Ozel}. On using this model} the estimated blackbody radius, R$_{\rm km}$, using the blackbody normalization (equation~\ref{eqn:1}),
\begin{equation}
 K= R^2_{\rm km}/D^2_{10}, 
 \label{eqn:1}
\end{equation}
turned out to be rather high $\sim$32 km, than the standard NS star blackbody radius, where, D$_{10}$ is the distance to the source in units of 10 kpc.

Then we replaced the blackbody component with disk blackbody component ({\tt diskbb}), assuming the accretion disk to be the source of the seed photons. Often, the disk contribution is generally stronger in the soft state. { Figure~\ref{fig:6spec} illustrates the representative spectra corresponding to FL I-A across the energy range of 0.7–20.0 keV using the model {\tt const*tbabs*(gauss+thComp*diskbb)}.} A faint hump in the residuals at $\sim$6.9 keV was observed (see Figure~\ref{fig:6spec}(c)), giving a hint of the presence of disk reflection feature. To fit this feature, we added a Gaussian ({\tt gauss}) component { to the continuum model ({\tt const*tbabs*(thComp*diskbb)})} with line energy fixed to 6.9 keV. The best-fit hydrogen column density ( N$_H$) agrees with the previously reported value, $\sim$1$\times$ 10$^{22}$ cm$^{-22}$ \citep{1989Penninx, 2010Guver, 2017Padilla}. The { color corrected} inner disk radius (R$_{\rm in(co)}$) has been calculated using the best-fit disk normalization constant (N$_{\rm dbb} $), following the relation:

\begin{equation}
\centering
R_{\rm in(co)}= \kappa^2(N_{\rm dbb}/cos\theta)^{1/2}*D_{10}
\label{eqn:2}
\end{equation}
where $\theta$, the inclination angle of the disk, is taken as 40$^{\circ}$ \citep{2015Degenaar}. { \citet{2015Degenaar} used {\it NuSTAR} observations to estimate the inclination angle range of the source between $\sim$ 30$^{\circ}$-40$^{\circ}$, constraining the R$_{\rm in}$ to be 7-10 $GM/c^2$}. And $\kappa$ ($\sim$1.7-2.0), is the color-hardening factor (ratio of color temperature to the effective temperature) \citep{1995Shimura}. In our calculation, we considered the color-correction factor as 1.7 \citep{1995Shimura} and $D_{10}$ as 0.4 kpc \citep{2010Guver}. It result R$_{\rm in}$ in the range of { $\sim$ 22.17$^{+2.57}_{-2.38}$--27.19$^{+2.03}_{-1.85}$} km, which is in close agreement with the range reported by \citet{2015Degenaar} of $\sim$11-30 km (considering $\theta$ as 40$^{\circ}$). However, \citet{2017Padilla} obtained it in range $\sim$15-45 km taking $\theta$ as 70$^{\circ}$. Using {\tt diskbb} model \citet{2002Marek}, estimated the R$_{\rm in}$ to be $\sim$ 30 km, in BS.  
Errors are computed at 90\% confidence level. To take into account cross-calibration uncertainties a constant was applied, which was fixed for LAXPC 20 and set free for other instruments. The unabsorbed flux and disk flux for each { FL} was estimated using the {\sc XSPEC} model {\tt cflux}. Figure~\ref{fig:flux} represents the how the spectral parameters are varying with the total unabsorbed flux and the disk flux.

In addition, we calculated the mass-accretion rate ($\dot{\rm M}$)  which was not well estimated previously. The equation \citep{2008Galloway, 2019Mondal} we used is as follows,
\begin{equation}
\label{eqn:mdot}
\begin{split}
\dot{m}=&\:6.7\times 10^{3}\left(\frac{F_{\rm p}c_\text{bol}}{10^{-9} \text{erg}\: \text{cm}^{-2}\: \text{s}^{-1}}\right) \left(\frac{d}{10 \:\text{kpc}}\right)^{2} \left(\frac{1.4 M_{\odot}}{M_\text{NS}}\right)\\
 &\times\left(\frac{z+1}{1.31}\right) \left(\frac{10\:\text{km}}{R_\text{NS}}\right) \text{g}\: \text{cm}^{-2}\: \text{s}^{-1},
 \end{split} 
\end{equation}

\begin{figure}
\centering
 \includegraphics[scale=0.50]{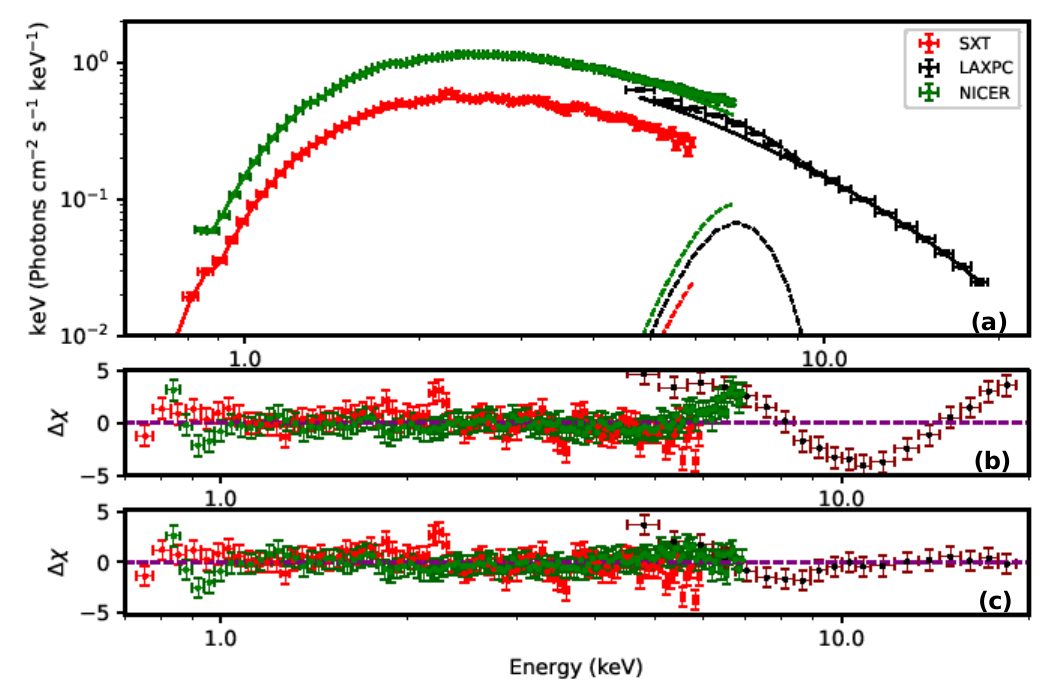}

 \caption{Time-resolved spectroscopy ({\it AstroSat} \& {\it NICER}): The representative spectrum and best-fit model spectrum over the energy range 0.7--20 keV for the segment I (see Figure~\ref{fig:lc_nicer}). The panels (a) and (c) show the spectra fitted using const*tbabs*(gauss+thComp*diskbb) model and its corresponding residuals, respectively, whereas the panel (b) shows the residual emission using the continuum model only.}
\label{fig:3spec}
\end{figure}

\begin{table*}
\centering
\setlength{\tabcolsep}{6.0pt}
	\caption{{ Time-resolved spectroscopy:} { Best-fit} spectral parameters { of the spectra} for simultaneous {\it AstroSat} (set-II) and {\it NICER} observation.}
	\label{tab:tab3}
	\begin{tabular}{@{}ccccc@{}} 
		\hline
		Spectral Component& Parameter & {Segment 1} & {Segment 2} & {Segment 3} \\
		
		\hline
{\tt Tbabs}&$N_{\rm H}$ [$10^{22}$~cm$^{-2}$] & 0.99 $^{+0.01}_{-0.01}$   &0.99  $^{+0.01}_{-0.01}$ &   0.98 $^{+0.01}_{-0.01}$\\
\hline
&&&&\\
{\tt gauss} &Line$_{\rm gauss}(keV) $&  6.9$^{*}$ & 6.9$^{*}$   &6.9$^{*}$\\
&&&&\\
&$\sigma (\rm keV)$&1.07 $^{+0.15}_{-0.14}$ & 1.16 $^{+0.14}_{-0.14}$ & 1.21  $^{+0.16}_{-0.15}$\\  
&&&&\\
 &N$_{\rm gauss}\times 10^{-2}$ &2.68 $^{+0.52}_{-0.47}$& 3.37 $^{+0.61}_{-0.55}$&  3.41 $^{+0.67}_{-0.61}$\\
\hline
&&&&\\
{\tt thComp} &$\Gamma $&  2.58 $^{+0.13}_{-0.11}$& 2.59  $^{+0.15}_{-0.13}$& 2.60 $^{+0.16}_{-0.13}$ \\ 
&&&&\\
  &  kT$_e$ (keV)& 3.88 $^{+0.64}_{-0.39}$ & 3.82  $^{+0.65}_{-0.42}$ & 3.88 $^{+0.72}_{-0.44}$\\ 
\hline
  &&&&\\
{\tt diskbb} &T$_{\rm in}$ (keV)&0.98 $^{+0.04}_{-0.04}$  & 1.08  $^{+0.05}_{-0.05}$&1.11  $^{+0.05}_{-0.05}$\\
&&&&\\
 &N$_{\rm dbb} $& 361 $^{+51}_{-45}$& 270 $^{+42}_{-34}$ & 256 $^{+38}_{-32}$\\ 
   &&&&\\
 
   \hline
&$\chi^2/\rm dof$ & 206.40/211  &  196.22/214 & 192.46/216 \\
            & &($\sim$ 0.98)&($\sim$ 0.92)&($\sim$ 0.89)\\
             &&&&\\
 \hline
&Inner disk radius (km)&8.66 $^{+0.59}_{-0.56}$& 7.49  $^{+0.56}_{-0.49}$& 7.29 $^{+0.52}_{-0.49}$ \\
&&&&\\
 \hline
&R$_{\rm in(co)}$&25.02$^{+1.71}_{-1.61}$& 21.65  $^{+1.61}_{-1.42}$& 21.07 $^{+1.50}_{-1.42}$ \\
\hline	
 & Mass accretion rate ($\dot{\rm M}$)    & 1.42 $^{+0.03}_{-0.03}$  & 1.55 $^{+0.04}_{-0.04}$ & 1.62 $^{+0.04}_{-0.03}$\\
   & (10$^{17}\rm gm/\rm s$)& & &\\
   \hline
   	
  \\
 {$^*$Frozen parameter.}\\
\multicolumn{5}{p{12cm}}{Notes: $N_{\rm H}$ is neutral hydrogen column density; $\sigma$ is Gaussian line width; $\Gamma$ is asymptotic power-law index; kT$_e$, T$_{\rm in}$ represents the electron and inner disk temperature respectively; $\chi^2/\rm dof$ is $\chi^2$ per degrees of freedom, the values below them are $\chi^2_{\rm red}$ value. R$_{\rm in(co)}$ (km) is color-corrected inner disk radius.}

\end{tabular}
\end{table*}

\begin{figure}
\centering
	\includegraphics[scale=0.70]{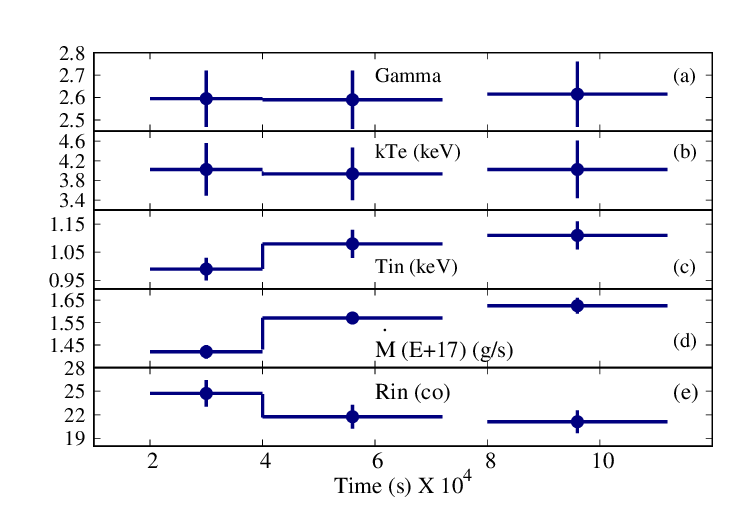}
    \caption{Variation of the spectral parameters ({ (a) Gamma ($\Gamma$)}, { (b) kT$_e$}, { (c) T$_{\rm in}$}, { (d) $\dot{\rm M}$}, and { (e) R$_{\rm in(co)}$)} vs time corresponding to the three segments shown in Figure~\ref{fig:lc_nicer}.}
    \label{fig:trs}
\end{figure}

\begin{figure}

 \includegraphics[scale=0.28,angle=-90]{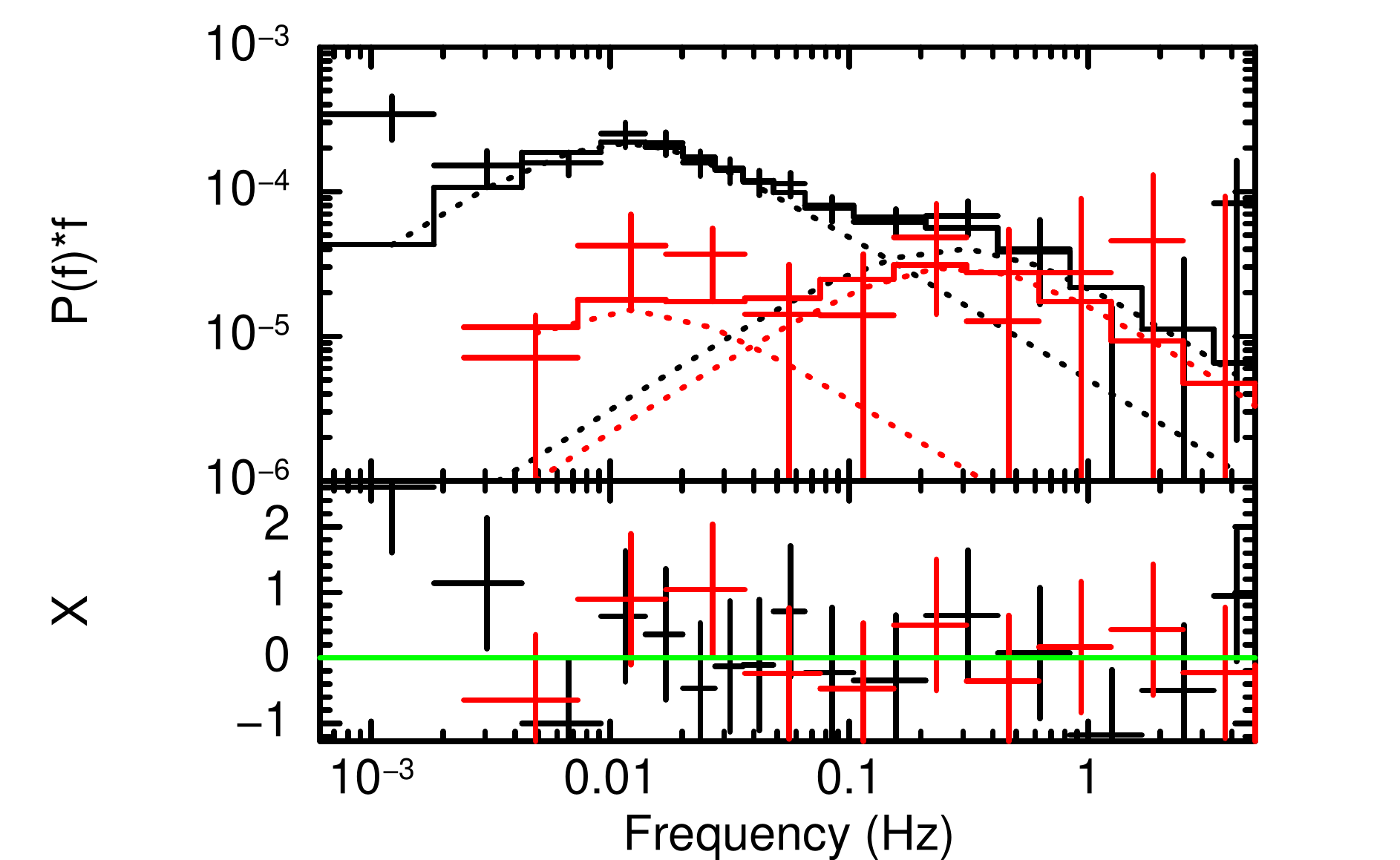}
 \includegraphics[scale=0.28,angle=-90]{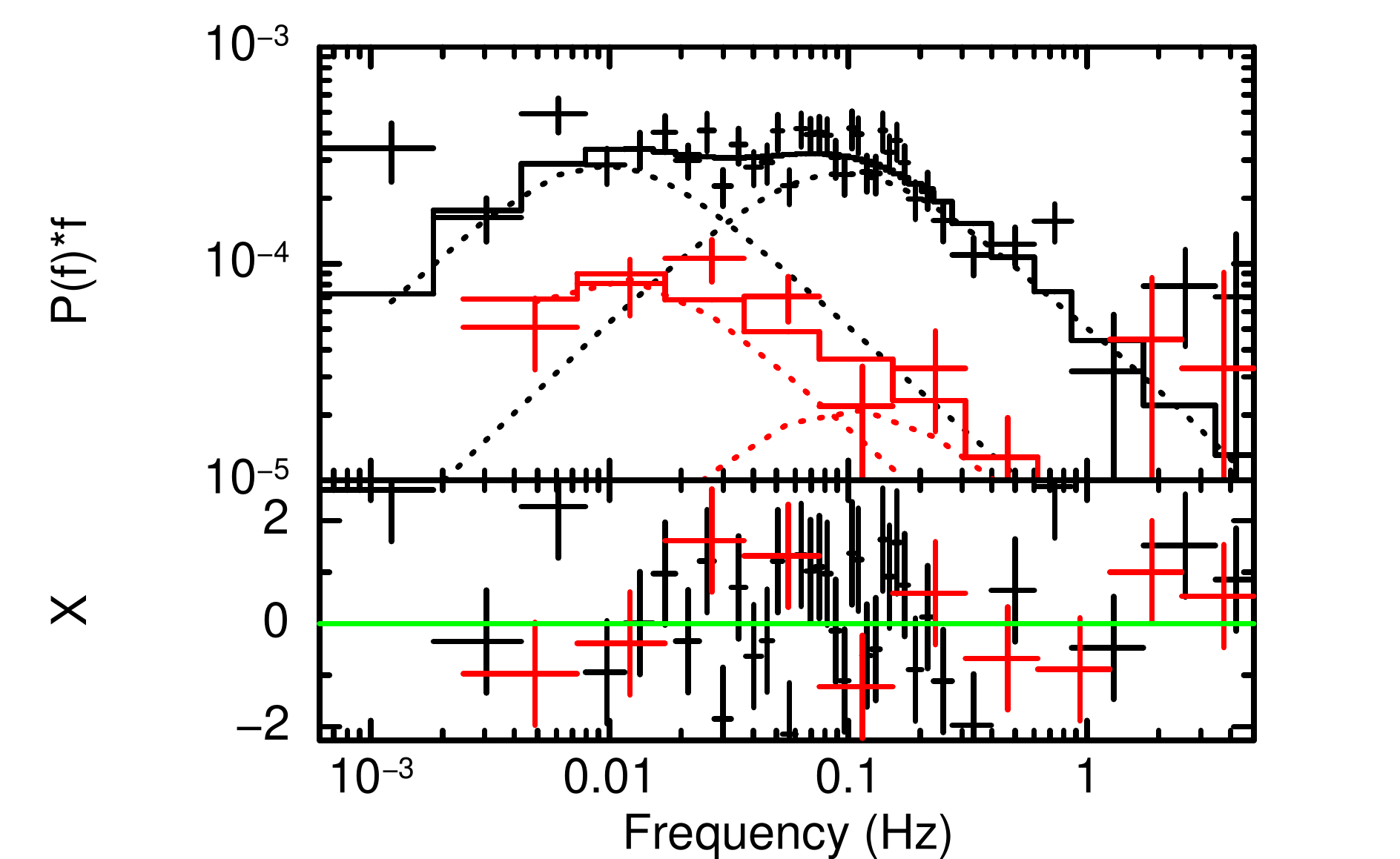}
 \includegraphics[scale=0.28,angle=-90]{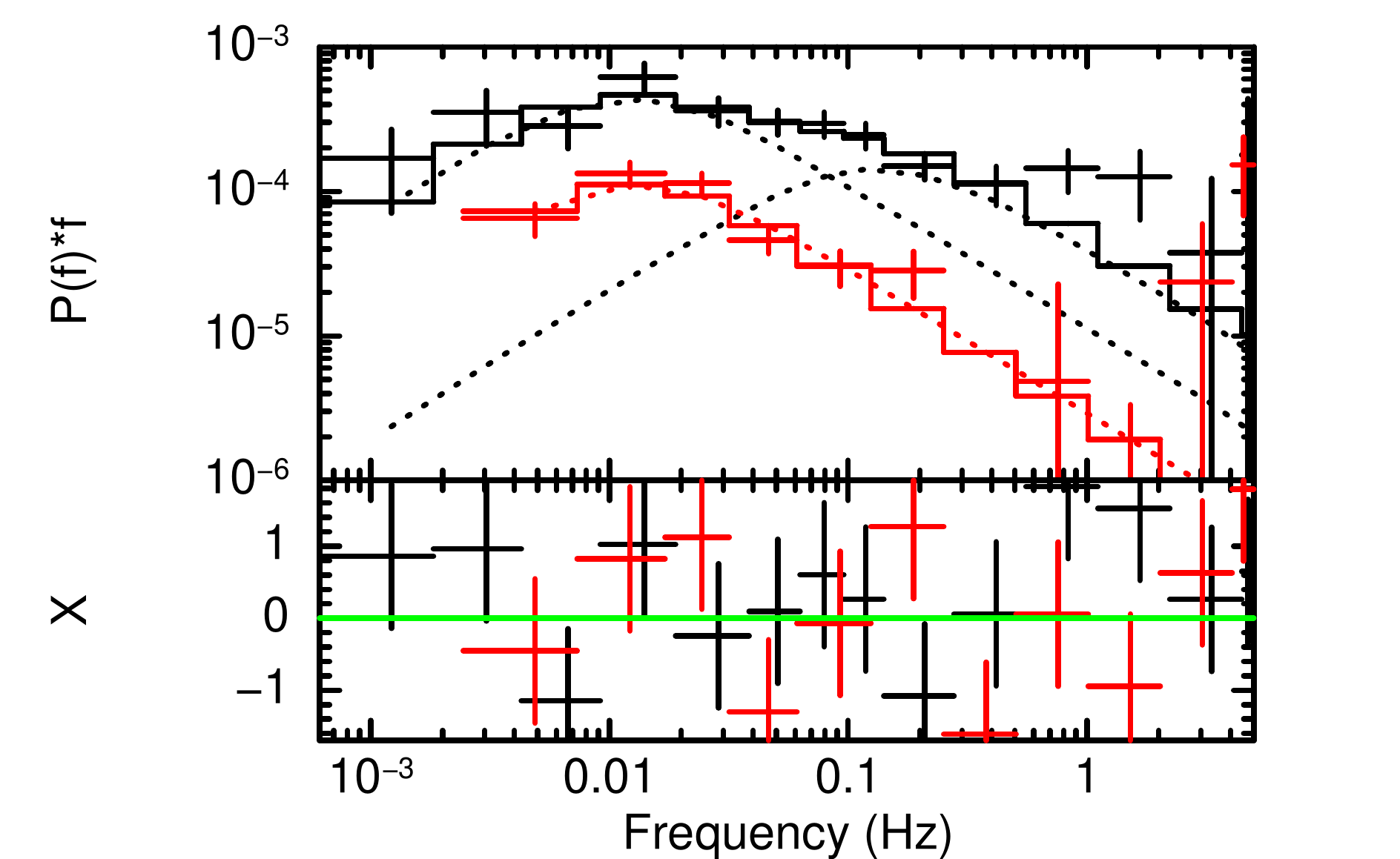}
\caption{ The PDSs and the corresponding model fits (see Figure~\ref{fig:lc_nicer}): segment 1 (top-panel), segment 2 (middle-panel), and segment 3 (bottom-panel), using the corresponding LAXPC (black) and {\it NICER} data (red). It is fitted using two zero-centered Lorentzians. }
\label{fig:pds1}
\end{figure}

where, $\dot{m}$ estimates the accretion rate per unit area of the source (i.e. $\dot{\rm M}$=$\dot{m}$4$\pi$$R_{\rm NS}$$^2$), $F_{\rm p}$ is the persistent flux, $c_{\rm bol}$ is the bolometric correction, adopted as $\sim 1.38$, which is standard for the nonpulsing sources \citep{2008Galloway} and; $d$, $M_{\rm NS}$ and $R_{\rm NS}$ are the distance, mass and radius of the source. The value of the factor $z+1$ is taken as 1.31, which is standard for typical NS \citep{2019Mondal}. Mass accretion rate is found to be greater in set-I in comparison to the set-II (see Table~\ref{tab:tab2}), which satisfies with the most of the reported observations \citep{1989Mitsuda, 1989Hasinger, 1994vander, 2002Marek}. However, a clear understanding for such state transition is yet to be found, since at different luminosities also such transitions have been observed for similar kinds of sources \citep{2000Bloser}. 


\subsection{Time--resolved Spectroscopy}
\subsubsection{Using the {\it AstroSat} and {\it NICER} simultaneous observation}

We proceed to conduct a time-resolved spectroscopy of the source data. The simultaneous observation of {\it AstroSat} (set-II) and {\it NICER} were split into three segments, as represented in Figure~\ref{fig:lc_nicer}. The 4.0–20.0 keV and 0.7–6.0 keV data of LAXPC and SXT have been used, respectively, with a systematic uncertainty of 3\%. The energy band considered for {\it NICER} is 0.8–7.0 keV, since below 0.8 keV the count rate was relatively low, and at energies greater than 10 keV the systematic rises to 40\%\footnote{\url{https://heasarc.gsfc.nasa.gov/docs/nicer/analysis_threads/spectrum-systematic-error/}}, a systematic of 2\% was added to its spectrum. Hence, time-resolved spectroscopy has been undertaken over the energy range of 0.7–20.0 keV.

We fitted the spectra for each of the three segments (Figure~\ref{fig:lc_nicer}) with the same model { combination} that was used for the flux-resolved spectroscopy i.e. {\tt const*tbabs*(gauss+thComp*diskbb)}. Figure~\ref{fig:3spec} shows the representative spectra for segment-I, and Table~{\ref{tab:tab3} lists all the best-fit model parameters with 90\% confidence level. Using this model, the estimated R$_{\rm in(co)}$ value using equation~\ref{eqn:2}, was found to be better constrained and in the range of { $\sim$21.07$^{+1.50}_{-1.42}$--25.02$^{+1.71}_{-1.61}$ km}, considering $\theta$, $D_{10}$, and $\kappa$ as 40$^{\circ}$ \citep{2015Degenaar}, 0.4 kpc \citep{2010Guver} and 1.7 \citep{1995Shimura} respectively. Figure~\ref{fig:trs}, represents how the spectral parameters are varying with the segments. { The gap between the second and third segments in Figure~\ref{fig:trs} exists to incorporate with all the data from LAXPC, SXT, and {\it NICER} when all the three instruments were functional.} Most of the parameters are consistent with being non-variable with the exception of T$_{\rm in}$, which is slightly increasing while, R$_{\rm in(co)}$ is slightly decreasing.

\section{ Timing analysis}
\label{sec:TA}
\subsection{Power density spectra}
\label{sub:pds}

In the PDSs generated from both LAXPC and {\it NICER} data, we don't detect any significant QPOs and hence we limit the analysis to the study of the broad-band noise, which can also provide clues to the nature of the accretion flow. We use the subroutine {\it laxpc\_find\_freqlag}\footnote{{\url{ http://astrosat-ssc.iucaa.in/laxpcData}}} of LAXPCsoftware and {\tt NICER\_RM\_Software}\footnote{Software available upon request} to generate the PDS of LAXPC and {\it NICER}, respectively and to estimate the energy-dependent time-lag and fractional rms. { For a non-paralyzable detector, the dead time corrected Poisson (white) noise level is given by $ <P(\nu)> = 2+4 \times\left(\frac{-1+\cos(2\pi \nu t_d) -(2\pi \nu \tau) \sin(2\pi \nu
t_d)} {2+(2\pi \nu \tau)^2 - 2 \cos(2\pi \nu t_d) + 2 (2\pi \nu \tau)
\sin(2\pi \nu t_d)}\right)$ \citep{1995Zhang}, where $t_d$ is dead time,
$R_T = R_{TO}/(1-R_{TO} t_d)$ is the dead time corrected count rate for the detector, and $R_{TO}$ is the observed total count rate. $\tau = 1/R_T$, where $\tau$ is average time interval between two successive events. In our analysis, each LAXPC PDS is dead time (42 $\mu$s) corrected and Poisson (white) noise subtracted using this method, a detailed explanation of which is provided in \citet{2016Yadav}, where the authors have shown that the high-frequency power spectrum from 100 Hz till even 50,000 Hz can be described by a simple dead time model with an effective dead time of 42.3 $\mu$s. Moreover, the PDSs have been adjusted to consider the background count rates \citep{2016Yadav}. Similarly, dead time is corrected, and Poisson noise is subtracted for the {\it NICER} PDSs, too. However, the dead time is ignorable, <1\%\footnote{\url{https://heasarc.gsfc.nasa.gov/docs/nicer/data_analysis/workshops/NICER-Workshop-QA-2021.pdf}}.}

For the LAXPC data, we first generated the PDS in the energy band of 3.0 to 20.0 keV for each segment, using a time resolution of 0.1 seconds and by dividing the lightcurve into segments of length 819.2 seconds. Similarly, for {\it NICER} data, we generated PDS in the energy range 0.5 to 10.0 keV using a time resolution of 0.1 seconds and with segments of length 204.8  seconds. The Nyquist frequency is 5 Hz for all the cases. Figure~\ref{fig:pds1} shows the LAXPC-{\it NICER} joint-fit PDS for all three segments using two zero-centered Lorentzians. The same re-binning process has been used to produce the PDS for both the LAXPC and {\it NICER}. The PDS has been re-binned with a minimum signal-to-noise ratio (SNR) of 5 in both the cases. Table~\ref{tab:tab4} lists the best-fit parameters and $\chi^2$/d.o.f for the joint-fitting of each segment. Due to the large uncertainties in the {\it NICER} data, we fitted the PDS with the same Lorentzians as used for LAXPC, but allowed for the normalizations of the two components to vary.

From the PDS plots, we cannot detect any profound signature of the presence of the QPO, instead, we detected another feature, a broad-structure called noise. Apart from the specific study of the QPOs, probing the variability in general also offers a tool to study the accretion flow around the central compact object. Some such earlier studies are, e.g., \citet{1999Psaltis,1999Wijnands,1999Wij, 2002Belloni, 2003Yu, 2004Vander, 2006van, 2010Altamirano, 2011Maccarone, 2013Belloni, 2016Motta, 2022Yang}.

\begin{table*}
\small
\centering
\setlength{\tabcolsep}{2.50pt}
	\caption{\label{tab:tab4}The best-fit model parameters of the fitted power density spectrum using two zero-centered Lorentzians. `–' represents that the parameter value has been kept the same as the { {\it AstroSat} parameters.}}
 	\begin{tabular}[c]{c c c c c c c c c}
 \\\hline \hline

Segment & Mission & & Lorentzian 1 & & &Lorentzian 2 && $\chi^2$/dof\\
&&$\nu$ (Hz)&$\sigma$ (mHz) &Norm $\times 10^{-3}$ & $\nu$ (Hz)&$\sigma$ (Hz)&Norm $\times 10^{-3}$&\\\hline

&{\it AstroSat} &  0$^{*}$ & 21.48 $^{\rm +5.86}_{\rm -5.00}$ & 0.69 $^{\rm +0.10}_{\rm -0.10}$ &0$^{*}$ & 0.52 $^{+0.62}_{ -0.32}$& 0.13$^{\rm +0.06}_{\rm -0.06}$& 
 \\
1&&&&&&&&17.94/20\\
&{\it NICER}& -- &   --  &0.05 $^{\rm +0.01}_{\rm -0.01}$ & -- & -- &<0.21& \\
\hline

&{\it {\it AstroSat} } &  0$^{*}$ & 17.91 $^{\rm +6.81}_{\rm -4.91}$ & 0.90$^{\rm +0.17}_{\rm -0.17}$ &0$^{*}$ & 0.18 $^{+0.04}_{ -0.03}$& 0.82 $^{\rm +0.12}_{\rm -0.12}$&
 \\
2&&&&&&&&69.37/41\\
&{\it NICER} &   --  & -- & 0.29$^{\rm +0.09}_{\rm -0.09}$  & -- &--&0.07$^{\rm +0.06}_{\rm -0.06}$&\\
\hline

&{\it {\it AstroSat} }&  0$^{*}$& 23.26$^{\rm +5.65}_{\rm -5.57}$  & 1.43 $^{\rm +0.57}_{\rm -0.51}$ & 0$^{*}$ & 0.26 $^{+2.45}_{ -0.14}$& 0.45 $^{\rm +0.19}_{\rm -0.30}$&
 \\
  3&&&&&&&&24.45/19\\
  &{\it NICER} &    --  & -- &  0.37$^{\rm +0.06}_{\rm -0.07}$  & -- &--& <0.04&
\\
\hline
		{$^*$Frozen parameter.}
\end{tabular} 
\end{table*}

\subsection{The energy-dependent temporal properties}
\label{sub:modelling}

We calculated the energy-dependent rms variation and time-lag for the observed broad-band features in order to get insight into the radiative processes responsible for the variability. These were computed over a frequency range $f -\Delta f/2$ to $f +\Delta f/2$. We choose $f = 19$ mHz and $\Delta f = 9$ mHz to represent the low frequency noise component for all segments. To represent the higher-frequency noise component, for segment 1 we choose $f = 0.25$ Hz, and for segments 2 \& 3 $f = 0.12$ Hz and $\Delta f $is $0.06$ Hz for all the segments. The results are shown in Figures~\ref{fig:rms1} and ~\ref{fig:rms2}. The phase lags are computed with respect to the reference energy band i.e. 3.0-4.0 keV for LAXPC and 0.8-2.0 keV for {\it NICER} by dividing the entire energy range into small intervals.

For all segments and PDS components, the phase lags for both {\it NICER} and LAXPC are consistent with zero (Table ~\ref{tab:tab5}). The fractional rms shows an increasing trend with energy and the {\it NICER} and LAXPC results are more or less consistent with each other. The fractional rms is <10\% (Figure ~\ref{fig:rms1} \& ~\ref{fig:rms2}) in every segment, which is expected since the source was in a soft state during the observation \citep{1990Belloni, 1995Nowak}. The combination of both {\it NICER} and LAXPC provides fractional rms in an energy range of 0.8-15 keV, which can be exploited to understand the radiative process responsible for the variation. The energy-dependence can be predicted using a model where one or two spectral parameters are varying coherently with a possibility of a phase lag between them \citep{2019Maqbool, 2020Garg}. In particular, for the Comptonization component represented by the XSPEC model {\tt thcomp}, it is more physical to use the optical depth ($\tau$) instead of the asymptotic power-law index ($\Gamma$) which are related by \citep{1996Zdziarski,1999Zycki},
\begin{equation}
  \Gamma = [9/4+(3m_ec^2)/(kT_e((\tau+3/2)^2-9/4))]^{1/2}-1/2
  \label{eq:gam}
\end{equation}
\noindent
where $kT_e$ is the electron temperature. Instead of the electron temperature, one may use the heating rate of the corona  ($\dot H$). The transformation of these spectral parameters and the technique to numerically estimate the variation of the spectrum caused by small linear variations of these parameters are described in detail in \citet{2020Garg}.

\citet{2020Garg} has proposed a more generic approach to describe the energy-dependent nature of the phase lags and fractional rms behaviours at a peaked noise as well as for the low frequency QPOs observed in the BH-LMXBs. In the study by \citet{2020Garg}, they showcase the effectiveness of their model in capturing the temporal dynamics of GRS 1915+105. Specifically they have shown that the observed milli seconds lag in the BH-LMXBs can be well explained considering the stochastic fluctuation propagation model which explains that the change in the disk spectral parameters due to any perturbation that has originated at the inner disk region can propagate to the corona causing the coronal parameters to change accordingly on a viscous or sound crossing time scale rather than the light crossing one as related to the Compton scattering lags. Furthermore, \citet{2022Garg} apply this model to the quasi-periodic variability observed in the source MAXI J1535–571, resulting in the determination of both $kT_{\rm in}$ and $\dot H$. The associated variation of $\dot H$ is also explored in their analysis. In their work, they have explicitly used the variation on different spectral parameters like heating rate rather than the electron temperature, optical depth, disk normalization, scattering fractions and inner disk temperature and computed the probable fractional rms variation using 
\begin{equation}
\delta F(E) = \sum_{j}\frac{\delta F (E)}{\delta \beta_j}\Delta \beta_{j}
  \label{eq:frms}
\end{equation}
\noindent
where F(E) is the time averaged spectrum on which the variations are to be imposed. The $\delta F(E)$ signifies the changes on the spectrum due to any change in the parameter that is $\delta$$\beta$.

\begin{figure*}
\centering
\includegraphics[width=0.7\linewidth, height=0.5\textwidth]{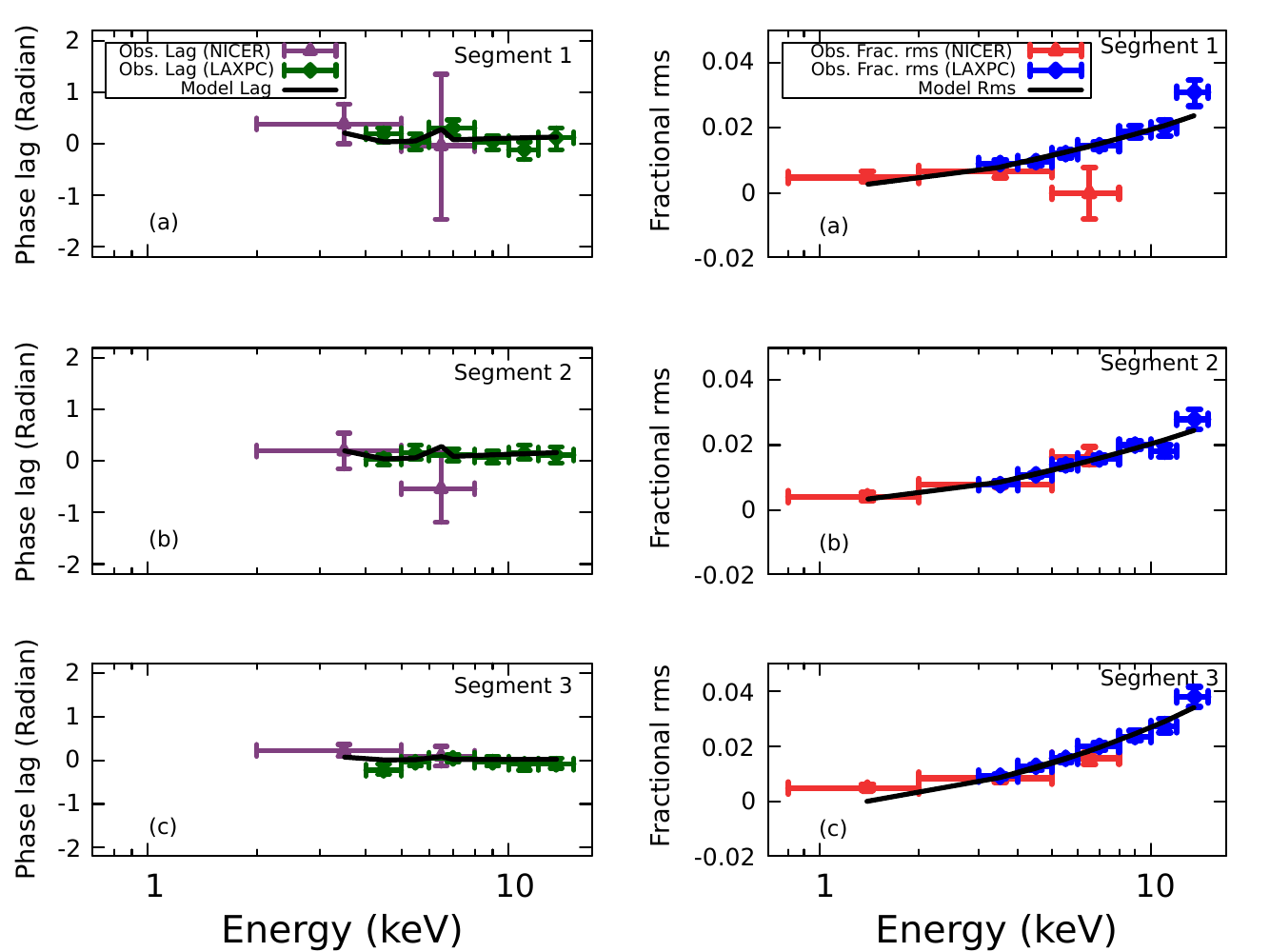}
\caption{ The phase lag (left-hand panel) and fractional rms (right-hand panel) as a function of photon energy of lower-frequency broad-band noise measured at frequency 19 mHz, with a group binning of 9 mHz for the three segments 1, 2, and 3 as shown in panel (a), (b), and (c). Seven LAXPC energy bands (3.0–4.0, 4.0–5.0, 5.0–6.0, 6.0–8.0, 8.0–10.0, 10.0–12.0, 12.0–15.0 keV) and three {\it NICER} energy bands (0.8–2.0, 2.0–5.0, 5.0–8.0 keV) are considered to calculate the lag of the lower-frequency broad-band feature, where 3.0-4.0 keV and 0.8-2.0 keV are the reference energy bands for {\it AstroSat}/LAXPC and {\it NICER}/XTI, respectively. It is to be noted that while doing the phase-lag modelling we had to keep the corresponding reference energy bands fixed, so in the phase-lag plot (left-hand panel), data related to that particular energy band is absent. The black solid curve represents the theoretically predicted model fit.}
\label{fig:rms1}
\end{figure*}


In our case, the physical parameters whose variation can lead us to the observed energy-dependent variability are the disk normalization, N$_{\rm dbb} $ (which represents the inner disk radius), the inner disk temperature $kT_{\rm in}$, the corona optical depth ($\tau$) and its heating rate ($\dot H$). First, the command {\tt laxpc\_make\_lagrms\_pha} is used to generate the spectrum files corresponding to the lag event files of the LAXPC and {\it NICER}. Then, we tested different combination of parameters with coherent variations, which would describe the energy-dependent fractional rms and phase-lags. Since fractional rms increases with energy, variation in the coronal heating rate $\dot H$ is required, but not sufficient to explain the observed dependence. A combination of variations in $\dot H$ and the disk normalization N$_{\rm dbb}$ also did not give a good fit with a $\chi^2_{\rm red} > 2$ for some of the segments. Then we consider the possibile variation of $\dot H$ and $\tau$ but that also couldn't justify the scenario, as we noted a $\chi^2_{\rm red} > 3$ for the last two segments. Hence, we moved to the next combination of variations in $\dot H$ and the inner disk temperature, $kT_{\rm in}$, which provided the best-fit as shown in Figures~\ref{fig:rms1} and~\ref{fig:rms2} and the parameter values are listed in Table~\ref{tab:tab5}. As expected since the energy-dependent phase lags for the segments are consistent with zero, the phase lags between $kT_{\rm in}$ and $\dot H$ are also consistent with zero. The variation of $kT_{\rm in}$ was found to be less than that of $\dot H$ (refer Table~\ref{tab:tab5}). In Figures~\ref{fig:rms1} and~\ref{fig:rms2}, the model, represented by a black line, likewise fits and satisfies the observed variability.

\begin{figure*}
\centering
\includegraphics[width=0.7\linewidth, height=0.5\textwidth]{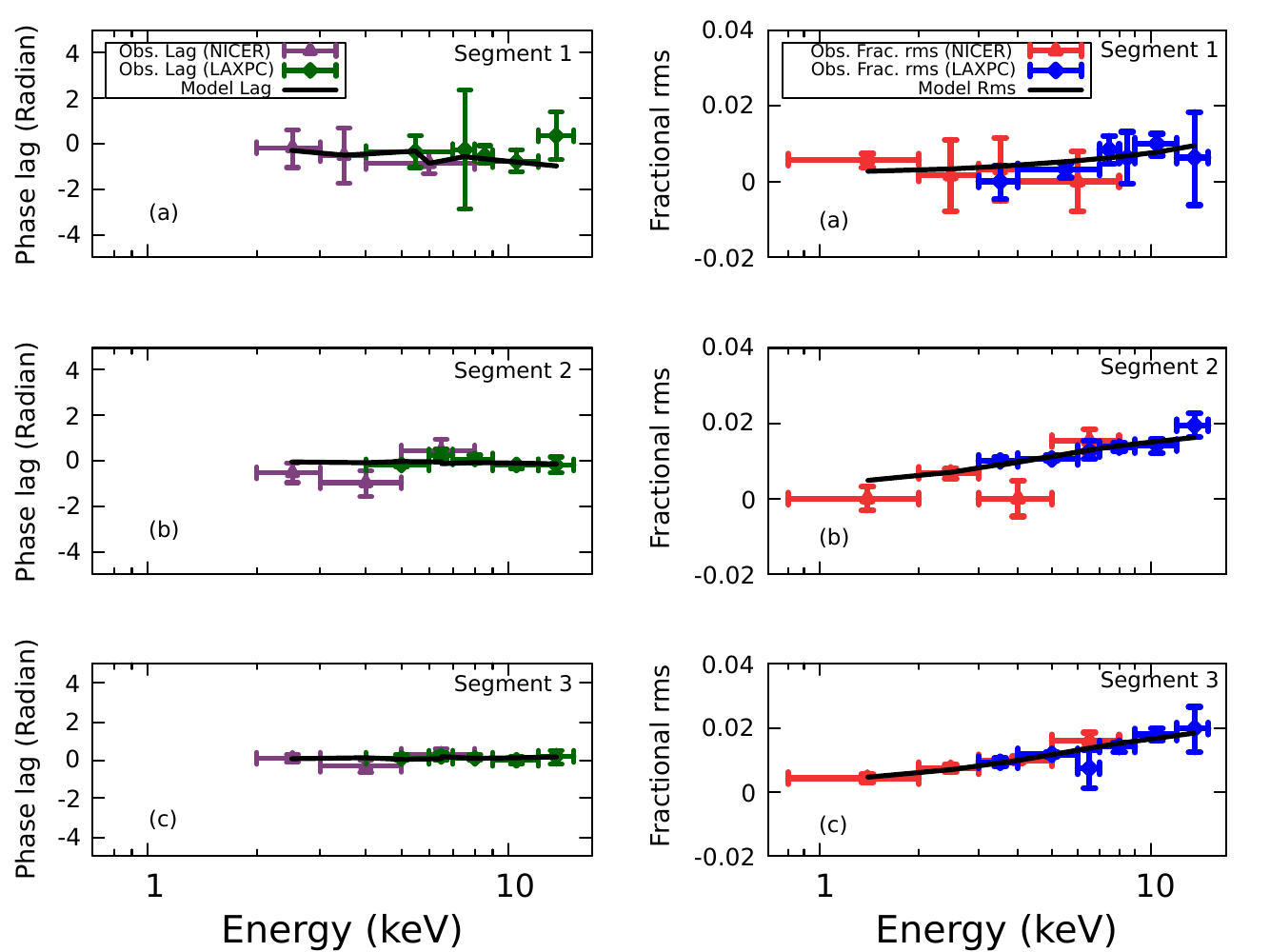}
\caption{The phase lag (left-hand panel) and fractional rms (right-hand panel) as a function of photon energy of higher-frequency broad-band noise measured at frequency 0.25 Hz, with a group binning of 0.06 Hz for the three segments 1, 2, and 3 as shown in panel (a), (b), and (c). Six
LAXPC energy bands (3.0–4.0, 4.0–5.0, 5.0–7.0, 7.0–9.0, 9.0–11.0,
11.0–15.0 keV) and four {\it NICER} energy bands (0.8–2.0, 2.0–3.0,
3.0–5.0, 5.0–8.0 keV) are considered to calculate the lags for the
higher-frequency broad-band feature, where 3.0-4.0 keV and 0.8-2.0 keV are the reference energy bands for LAXPC/{\it AstroSat} and XTI/{\it NICER}, respectively. It is to be noted that while doing the phase-lag modelling we had to keep the corresponding
reference energy bands fixed, so in the phase-lag plot (left-hand panel), data related to that particular energy band is absent. The black solid curve represents the theoretically predicted model fit.}

\label{fig:rms2}
\end{figure*}

\begin{table*}
\centering
	\caption{\label{tab:tab5} The Best-fit model parameters that describe the observed phase lag and fractional rms variation for the lower and higher frequency broad-band component.}
 	\begin{tabular}[c]{c c c c c c c }
\\\hline \hline
Broad-band component & Name of the segment & Varying Parameters & $\delta$\rm kT$_{\rm in}$ (\%)  & $\delta$$\dot H$ (\%) &Phase lag & $X^2$/dof\\\hline 

Lower-frequency& 1 &  $\delta $\rm kT$_{\rm in}$ $\rm w.r.t. \delta$$\dot H$  & $0.49 \pm 0.03$ & $4.29\pm 0.70$ & $-0.08\pm 0.10$& 17.71/15
 \\
&&&&&&\\
&2 & $\delta $\rm kT$_{\rm in}$ $\rm w.r.t. \delta$$\dot H$ & $0.56\pm 0.03$ & $3.18\pm 0.40$ & $ -0.11\pm 0.09$ &8.92/15
 \\
&&&&&&\\
&3 & $\delta $\rm kT$_{\rm in}$ $ \rm w.r.t. \delta$$\dot H$ & $0.72 \pm 0.03$ & $6.92\pm 0.70$ & $-0.01\pm 0.03 $& 12.33/14
 \\ 
\hline
Higher-frequency& 1 &  $\delta$\rm kT$_{\rm in}$  $\rm w.r.t. \delta$$\dot H$  & $0.20 \pm 0.07$ & $1.84\pm 1.00$ & $0.69\pm 0.58 $& 8.39/15
 \\
&&&&&&\\
&2 & $\delta $\rm kT$_{\rm in}$ $\rm w.r.t. \delta$$\dot H$ & $0.46 \pm 0.03$ & $2.33\pm 0.64$ & $-0.14\pm 0.24 $& 17.43/14\\
&&&&&&\\
&3 & $\delta $\rm kT$_{\rm in}$  $\rm w.r.t. \delta$$\dot H$ & $0.51 \pm 0.04$ & $3.07\pm 0.75$ & $-0.16\pm 0.20 $& 7.92/15\\
\hline
\end{tabular} 
\end{table*}

\section{discussion \& Summary}
\label{sec:disc}
We performed broad-band spectral and timing studies of the NS binary candidate 4U 1608--52.  We used two  data sets of {\it AstroSat} observations (data set-I and II) and two data sets of {\it NICER} observations, the same observation date as {\it AstroSat} data set-II. These data set allowed for the study of the spectral evolution of the source. The nearly simultaneous {\it AstroSat} and {\it NICER} data provided rapid timing information in the energy band of 0.8-15.0 keV.

Using the two {\it AstroSat} data sets, we performed flux-resolved spectroscopy to analyze the behavior of the source with increasing X-ray intensity. We carried out time-resolved spectroscopy for the {\it AstroSat}/{\it NICER} simultaneous observation by creating time segment containing  data from LAXPC/SXT and {\it NICER} (excluding the burst from the {\it NICER} observation). { For the spectral fitting, we have not used two separate components for the disk emission and the Comptonization. Instead, we used the convolutional model ThComp acting on the disk emission. Specifically, this is represented as ThComp*diskbb. Here, the parameters of Thcomp are the temperature, the photon index, and the covering fraction. When the covering fraction is zero, the output spectrum is the same as that of disk emission. Even when the covering fraction is not zero, the norm of disk emission relates to the one we see if the Comptonizing component was not there. Thus, the radius inferred from the disk emission in this formalism is the actual one. However, we have assumed here that the covering fraction is unity since spectral fitting cannot constrain it. This assumption may indeed change the reported disk radius. Further, our $R_{\rm in}$ value calculated depends the $\kappa$, $D_{10}$, and $\theta$ values, which we have considered from the literature. Thus, our measured $R_{\rm in}$ value is limited by the assumed $\kappa$, $D_{10}$, and $\theta$ values. The measured $T_{\rm in}$ is limited by the data quality and the model's fit to the data.}

 From the broad-band spectral analysis in the 0.7–20.0 keV energy band (Table~~\ref{tab:tab2}), the source was found to be in a soft spectral state ($\Gamma >$ 2) or in BS, during both the {\it AstroSat} observations. Figure~\ref{fig:HID_astrosat} reveals that during the first observation, it was in the UBS with higher X-ray luminosity and a slight upward-curvy pattern, while in the second one it was in the LBS, with a lower X-ray luminosity. 

 We estimated the total unabsorbed and disk flux for different { FLs}.
  The total unabsorbed flux was about $\sim  2.2 \times 10^{-8}$ ergs/s for the UBS which was about a factor of two higher than the flux in the LBS, $\sim 1.2 \times 10^{-8}$ ergs/sec. This translates to an accretion rate of $\sim 3.1 \times 10^{17}$ g/s and $\sim 1.8 \times 10^{17}$ g/s for the two states, which is consistent with  many of the earlier results for  atoll sources  \citep{1989Mitsuda, 1989Hasinger, 1994vander, 2002Marek}. The disk flux also undergoes a change between the two branches by decreasing from $\sim 1.3$ to $\sim 0.8 \times 10^{-8}$ ergs/sec. However, the color corrected inner disk radius remained nearly unchanged in all { FLs}, $\sim 23.78^{\rm +3.40}_{\rm -1.82}$ km and the decrease in disk flux from the upper to the LBS can be ascribed to a decrease in inner disk temperature. This implies that the overall geometry of the system (as in a disk truncated at a fixed  radius) did not vary, and the variation between the UBS and LBS can be primarily associated with the accretion rate. For the Comptonizing corona, the photon index $\Gamma$ was higher in the LBS as compared to the upper one. The electron temperature seemed to be higher for the lower state ($\sim 4$ keV) compared to the upper one ($\sim 3$ keV), but the error bars were too large to make any concrete statement. Within the LBS, $\Gamma$ was found to anti-correlate with the total flux, and thus the source became harder with increasing flux. Time resolved spectroscopy for the LBS using {\it AstroSat} and {\it NICER} data, gave consistent results with those obtained from the flux-resolved {\it AstroSat} data alone. In particular, the color corrected inner disk radii was found to have the same value of $\sim 22.58^{\rm +2.44}_{\rm -1.51}$ km. 
  
The nearly simultaneous {\it AstroSat}/LAXPC and {\it NICER} data provided the opportunity to study the rapid variability of the source in a unique energy band of 0.8-15.0 keV, in the LBS. The PDS for the different time segments revealed two broad features which were consistent for both LAXPC and {\it NICER}. For the two components, we estimated the fractional rms and phase-lags as a function of energy. While, the phase-lags were consistent with zero, the fractional rms showed a increasing trend with energy. The energy-dependence was then modelled using the formalism presented by  \citet{2020Garg}, where two physical spectral parameters vary coherently with a phase lag between them. We find that the data is well represented by a scenario, where there is variation of the inner disk temperature and the heating rate of the corona. Thus, the temporal modelling is also consistent with the spectral interpretation, where the inner disk radius does not vary and the rapid variability (as well as the longer timescale movement of the source along the banana branch) can be attributed to accretion rate variations, which changes the inner disk temperature and the heating rate of the corona.  

Our analysis is subjective to the limited simultaneous coverage of the source by {\it AstroSat} and {\it NICER}, which was limited to the lower banana branch. The results highlight the need for more extensive simultaneous observations, which will allow for a more detailed understanding of the spectral and timing properties of these sources.

\section*{ACKNOWLEDGMENTS}

This work has utilised the softwares provided by the High Energy Astrophysics Science Archive Research Centre (HEASARC). This work has used the LAXPC and SXT instruments data of {\it AstroSat} mission of ISRO. We thank the LAXPC Payload Operation Center (POC) and the SXT POC at TIFR, Mumbai, for providing the data via the ISSDC data archive and the necessary software tools. We thank XTI/{\it NICER} team too, for providing the analysis guidelines in their website. S.B. acknowledges the periodic visits to Inter-University Centre for Astronomy and Astrophysics (IUCAA), Pune, to carry out a significant portion of the work. Along with acknowledging the IUCAA visiting associateship programme, B.S. also thanks IUCAA for their hospitality throughout his visits, where part of this work was completed. This research has made use of MAXI data provided by RIKEN, JAXA and the MAXI team \citep{2009Matsuoka}. Also, we acknowledge the use of public data from the Swift data archive. S.B. acknowledges the financial support by the DST-INSPIRE (Reg. No.: DST/INSPIRE Fellowship/[IF220164]) fellowship by the Government of India. The authors, B.S. and A.N. acknowledge the financial support of ISRO under {\it AstroSat} archival Data utilisation program (No.DS-2B-13013(2)/2/2019-Sec.2). A.B. is supported by an INSPIRE Faculty grant (DST/INSPIRE/04/2018/001265) by the Departmentof Science and Technology, Govt. of India and also acknowledges the financial support of ISRO under {\it AstroSat} archival Data utilisation program (No.DS-2B-13013(2)/4/2019-Sec. 2). A.B. also acknowledges SERB (SB/SRS/2022-23/124/PS) for financial support. { Finally, we thank the anonymous referee for providing useful comments and suggestions that considerably improved the paper’s content.}

\facilities{ADS, HEASARC, {\it AstroSat} (\citealt{2014Singh}), {\it NICER}}
\software{\textsc{HEAsoft} (v. 6.31.1), \textsc{XSPEC} (12.13.0c;  \citealt{1996Arnaud})}


\begin{appendix}
\section{LAXPC background spectra}
\renewcommand\thefigure{\thesection\arabic{figure}} 

\begin{figure}[ht]
    \centering
    \includegraphics[scale=0.3,angle=-90]{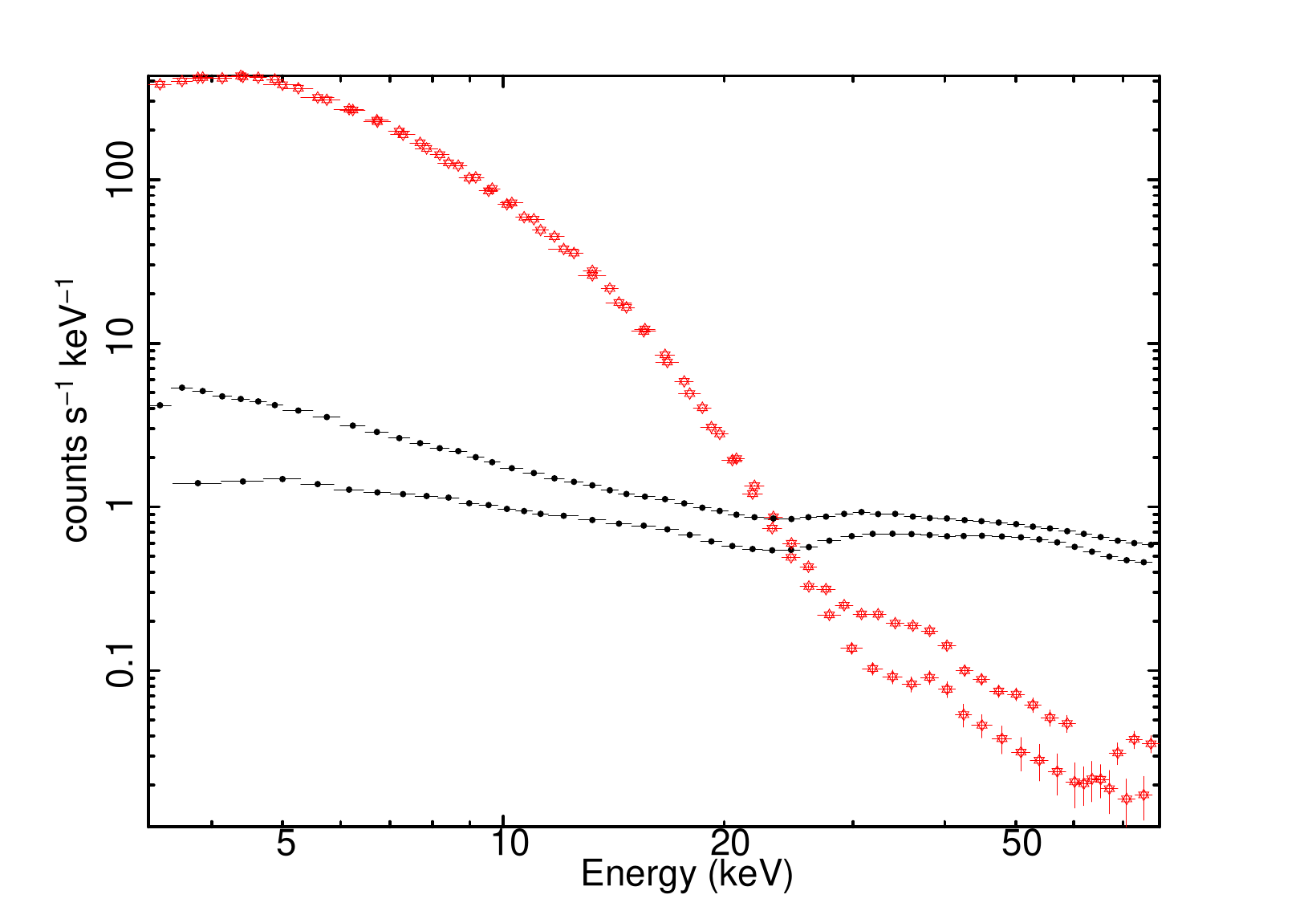}
        \includegraphics[scale=0.3, angle=-90]{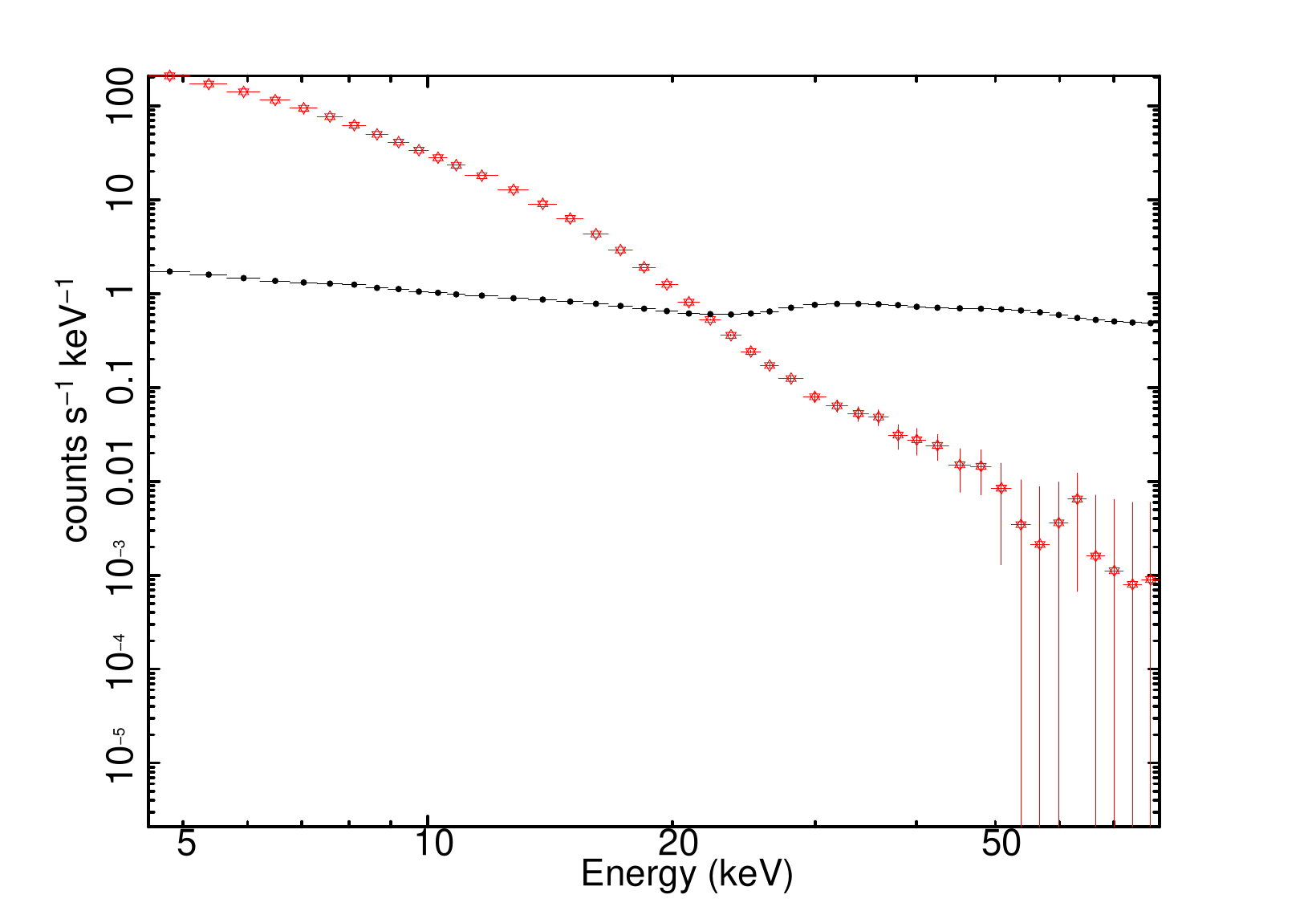}
    \caption{\label{fig:apdx1}Left: Plot showing LAXPC 10 and LAXPC 20 source with their background spectra for the set-I. Right: Plot showing LAXPC 20 source and background spectra simultaneously for set-II. In both the plots, it can be seen that there is a background dominance at higher energies. }
\end{figure}

\end{appendix}


\bibliography{draft1_apj_review_2}{}

\begin{thebibliography}{}
\expandafter\ifx\csname natexlab\endcsname\relax\def\natexlab#1{#1}\fi
\providecommand{\url}[1]{\href{#1}{#1}}
\providecommand{\dodoi}[1]{doi:~\href{http://doi.org/#1}{\nolinkurl{#1}}}
\providecommand{\doeprint}[1]{\href{http://ascl.net/#1}{\nolinkurl{http://ascl.net/#1}}}
\providecommand{\doarXiv}[1]{\href{https://arxiv.org/abs/#1}{\nolinkurl{https://arxiv.org/abs/#1}}}

\bibitem[Agrawal(2006)]{2006Agrawal} Agrawal, P.~C.\ 2006, Advances in Space Research, 38, 2989, \dodoi{doi:10.1016/j.asr.2006.03.038}

\bibitem[Agrawal et al.(2017)]{2017Agrawal} Agrawal, P.~C., Yadav, J.~S., Antia, H.~M., et al.\ 2017, Journal of Astrophysics and Astronomy, 38, 30, \dodoi{doi:10.1007/s12036-017-9451-z}

\bibitem[Altamirano et al.(2010)]{2010Altamirano} Altamirano, D., Linares, M., Patruno, A., et al.\ 2010, \mnras, 401, 223, \dodoi{doi:10.1111/j.1365-2966.2009.15627.x}

\bibitem[Antia et al.(2017)]{2017Antia} Antia, H.~M., Yadav, J.~S., Agrawal, P.~C., et al.\ 2017, \apjs, 231, 10, \dodoi{ doi:10.3847/1538-4365/aa7a0e}

\bibitem[Antia et al.(2021)]{2021Antia} Antia, H.~M., Agrawal, P.~C., Dedhia, D., et al.\ 2021, Journal of Astrophysics and Astronomy, 42, 32, \dodoi{ doi:10.1007/s12036-021-09712-8}

\bibitem[Armas Padilla et al.(2017)]{2017Padilla} Armas Padilla, M., Ueda, Y., Hori, T., et al.\ 2017, \mnras, 467, 290, \dodoi{ doi:10.1093/mnras/stx020}
\bibitem[Arnaud(1996)]{1996Arnaud} Arnaud, K.~A.\ 1996, Astronomical Data Analysis Software and Systems V, 101, 17

\bibitem[Arzoumanian et al.(2014)]{2014Arzoumanian} Arzoumanian, Z., Gendreau, K.~C., Baker, C.~L., et al.\ 2014, \procspie, 9144, 914420, \dodoi{ doi:10.1117/12.2056811}

\bibitem[Asai et al.(2012)]{2012Asai} Asai, K., Matsuoka, M., Mihara, T., et al.\ 2012, \pasj, 64, 128, \dodoi{ doi:10.1093/pasj/64.6.128}


\bibitem[Barret(2001)]{2001Barret} Barret, D.\ 2001, Advances in Space Research, 28, 307, \dodoi{ doi:10.1016/S0273-1177(01)00414-8}

\bibitem[Barret(2013)]{2013Barret} Barret, D.\ 2013, \apj, 770, 9, \dodoi{ doi:10.1088/0004-637X/770/1/9}

\bibitem[Barua et al.(2020)]{2020Barua} Barua, S., Jithesh, V., Misra, R., et al.\ 2020, \mnras, 492, 3041, \dodoi{ doi:10.1093/mnras/staa067}

\bibitem[Belian et al.(1976)]{1976Belian} Belian, R.~D., Conner, J.~P., \& Evans, W.~D.\ 1976, \apjl, 206, L135, \dodoi{ doi:10.1086/182151
}
\bibitem[Bellavita et al.(2022)]{2022Bellavita} Bellavita, C., Garc{\'\i}a, F., M{\'e}ndez, M., et al.\ 2022, \mnras, 515, 2099, \dodoi{ doi:10.1093/mnras/stac1922}

\bibitem[Belloni \& Hasinger(1990)]{1990Belloni} Belloni, T. \& Hasinger, G.\ 1990, \aap, 230, 103

\bibitem[Belloni et al.(2002)]{2002Belloni} Belloni, T., Psaltis, D., \& van der Klis, M.\ 2002, \apj, 572, 392, \dodoi{ doi:10.1086/340290}

\bibitem[Belloni \& Altamirano(2013)]{2013Belloni}Belloni, T.~M. \& Altamirano, D.\ 2013, \mnras, 432, 19, \dodoi{doi:10.1093/mnras/stt285}


\bibitem[Beri et al.(2019)]{2019Beri} Beri, A., Paul, B., Yadav, J.~S., et al.\ 2019, \mnras, 482, 4397, \dodoi{ doi:10.1093/mnras/sty2975}

\bibitem[Beri et al.(2021)]{2021Beri} Beri, A., Naik, S., Singh, K.~P., et al.\ 2021, \mnras, 500, 565, \dodoi{ doi:10.1093/mnras/staa3254}

\bibitem[Beri et al.(2023)]{2023Beri} Beri, A., Sharma, R., Roy, P., et al.\ 2023, \mnras, 521, 5904, \dodoi{ doi:10.1093/mnras/stad902}

\bibitem[Bloser et al.(2000)]{2000Bloser} Bloser, P.~F., Grindlay, J.~E., Barret, D., et al.\ 2000, \apj, 542, 989, \dodoi{ doi:10.1086/317035}

\bibitem[Campana et al.(1998)]{1998Campana} Campana, S., Colpi, M., Mereghetti, S., et al.\ 1998, \aapr, 8, 279, \dodoi{doi:10.1007/s001590050012}

\bibitem[Chen et al.(2019)]{2019Chen} Chen, Y.~P., Zhang, S., Zhang, S.~N., et al.\ 2019, Journal of High Energy Astrophysics, 24, 23, \dodoi{ doi:10.1016/j.jheap.2019.09.001}

\bibitem[Chen et al.(2022)]{2022Chen} Chen, Y.-P., Zhang, S., Ji, L., et al.\ 2022, \apj, 936, 46, \dodoi{ doi:10.3847/1538-4357/ac87a0}

\bibitem[Church et al.(2006)]{2006Church} Church, M.~J., Halai, G.~S., \& Ba{\l}uci{\'n}ska-Church, M.\ 2006, \aap, 460, 233, \dodoi{doi:10.1051/0004-6361:20065035}

\bibitem[Church et al.(2014)]{2014Church} Church, M.~J., Gibiec, A., \& Ba{\l}uci{\'n}ska-Church, M.\ 2014, \mnras, 438, 2784, \dodoi{ doi:10.1093/mnras/stt2364}

\bibitem[de Avellar et al.(2013)]{2013Avellar} de Avellar, M.~G.~B., M{\'e}ndez, M., Sanna, A., et al.\ 2013, \mnras, 433, 3453, \dodoi{ doi:10.1093/mnras/stt1001}

\bibitem[Degenaar et al.(2015)]{2015Degenaar} Degenaar, N., Miller, J.~M., Chakrabarty, D., et al.\ 2015, \mnras, 451, L85, \dodoi{ doi:10.1093/mnrasl/slv072}

\bibitem[Degenaar et al.(2016)]{2016Degenaar} Degenaar, N., Koljonen, K.~I.~I., Chakrabarty, D., et al.\ 2016, \mnras, 456, 4256, \dodoi{ doi:10.1093/mnras/stv2965}

\bibitem[Di Salvo et al.(2001)]{2001Disalvo} Di Salvo, T., M{\'e}ndez, M., van der Klis, M., et al.\ 2001, \apj, 546, 1107, \dodoi{ doi:10.1086/318278}

\bibitem[Done et al.(2007)]{2007Done} Done, C., Gierli{\'n}ski, M., \& Kubota, A.\ 2007, \aapr, 15, 1, \dodoi{ doi:10.1007/s00159-007-0006-1}

\bibitem[Falanga et al.(2006)]{2006Falanga} Falanga, M., G{\"o}tz, D., Goldoni, P., et al.\ 2006, \aap, 458, 21, \dodoi{ doi:10.1051/0004-6361:20065297}


\bibitem[Galloway et al.(2008)]{2008Galloway} Galloway, D.~K., Muno, M.~P., Hartman, J.~M., et al.\ 2008, \apjs, 179, 360, \dodoi{ doi:10.1086/592044}

\bibitem[Garg et al.(2020)]{2020Garg} Garg, A., Misra, R., \& Sen, S.\ 2020, \mnras, 498, 2757, \dodoi{ doi:10.1093/mnras/staa2506}

\bibitem[Garg et al.(2022)]{2022Garg} Garg, A., Misra, R., \& Sen, S.\ 2022, \mnras, 514, 3285, \dodoi{doi:10.1093/mnras/stac1490}

\bibitem[Gendreau et al.(2012)]{2012Gendreau} Gendreau, K.~C., Arzoumanian, Z., \& Okajima, T.\ 2012, \procspie, 8443, 844313, \dodoi{ doi:10.1117/12.926396}

\bibitem[Gierli{\'n}ski \& Done(2002)]{2002Marek} Gierli{\'n}ski, M. \& Done, C.\ 2002, \mnras, 337, 1373, \dodoi{ doi:10.1046/j.1365-8711.2002.06009.x}

\bibitem[Gilfanov et al.(2003)]{2003Gilfanov} Gilfanov, M., Revnivtsev, M., \& Molkov, S.\ 2003, \aap, 410, 217, \dodoi{ doi:10.1051/0004-6361:20031141}

\bibitem[G{\"u}ver et al.(2010)]{2010Guver} G{\"u}ver, T., {\"O}zel, F., Cabrera-Lavers, A., et al.\ 2010, \apj, 712, 964, \dodoi{ doi:10.1088/0004-637X/712/2/964}

\bibitem[G{\"u}ver et al.(2021)]{2021Guver} G{\"u}ver, T., Boztepe, T., G{\"o}{\u{g}}{\"u}{\c{s}}, E., et al.\ 2021, \apj, 910, 37, \dodoi{ doi:10.3847/1538-4357/abe1ae}

\bibitem[Hanawa(1989)]{1989Hanawa} Hanawa, T.\ 1989, \apj, 341, 948, \dodoi{ doi:10.1086/167553}

\bibitem[Hartman et al.(2003)]{2003Hartman} Hartman, J.~M., Chakrabarty, D., Galloway, D.~K., et al.\ 2003, AAS/High Energy Astrophysics Division \#7

\bibitem[Hasinger \& van der Klis(1989)]{1989Hasinger} Hasinger, G. \& van der Klis, M.\ 1989, \aap, 225, 79

\bibitem[Hasinger et al.(1990)]{1990Hasinger} Hasinger, G., van der Klis, M., Ebisawa, K., et al.\ 1990, \aap, 235, 131

\bibitem[Homan et al.(2002)]{2002Homan} Homan, J., van der Klis, M., Jonker, P.~G., et al.\ 2002, \apj, 568, 878, \dodoi{ doi:10.1086/339057}

\bibitem[Homan et al.(2010)]{2010Homan} Homan, J., van der Klis, M., Fridriksson, J.~K., et al.\ 2010, \apj, 719, 201, \dodoi{ doi:10.1088/0004-637X/719/1/201}

\bibitem[Jaisawal et al.(2019)]{2019Jaisawal} Jaisawal, G.~K., Chenevez, J., Bult, P., et al.\ 2019, \apj, 883, 61, \dodoi{ doi:10.3847/1538-4357/ab3a37}

\bibitem[Ji et al.(2014)]{2014Ji} Ji, L., Zhang, S., Chen, Y., et al.\ 2014, \apjl, 791, L39 , \dodoi{doi:10.1088/2041-8205/791/2/L39}

\bibitem[Jithesh et al.(2019)]{2019Jithesh} Jithesh, V., Maqbool, B., Misra, R., et al.\ 2019, \apj, 887, 101, \dodoi{ doi:10.3847/1538-4357/ab4f6a}

\bibitem[Kaastra \& Bleeker(2016)]{2016Kaastra} Kaastra, J.~S. \& Bleeker, J.~A.~M.\ 2016, \aap, 587, A151, \dodoi{ doi:10.1051/0004-6361/201527395}

\bibitem[Kotov et al.(2001)]{2001Kotov} Kotov, O., Churazov, E., \& Gilfanov, M.\ 2001, \mnras, 327, 799, \dodoi{doi:10.1046/j.1365-8711.2001.04769.x}

\bibitem[Kuulkers et al.(2002)]{2002Kuulkers} Kuulkers, E., Homan, J., van der Klis, M., et al.\ 2002, \aap, 382, 947, \dodoi{doi:10.1051/0004-6361:20011656}

\bibitem[Kumar \& Misra(2014)]{2014Kumar} Kumar, N. \& Misra, R.\ 2014, \mnras, 445, 2818, \dodoi{ doi:10.1093/mnras/stu1946}

\bibitem[Lee et al.(2001)]{2001Lee} Lee, H.~C., Misra, R., \& Taam, R.~E.\ 2001, \apjl, 549, L229, \dodoi{ doi:10.1086/319171}

\bibitem[Lewin et al.(1987)]{1987Lewin} Lewin, W.~H.~G., Penninx, W., van Paradijs, J., et al.\ 1987, \apj, 319, 893, \dodoi{ doi:10.1086/165506}

\bibitem[Lewin \& van der Klis(2006)]{2006Lewin} Lewin, W.~H.~G. \& van der Klis, M.\ 2006, Compact stellar X-ray sources. Cambridge Astrophysics Series, No. 39. Cambridge, UK: Cambridge University Press, ISBN 978-0-521-82659-4, ISBN 0-521-82659-4, \dodoi{doi: 10.2277/0521826594, 2006, XV+690 pp}.

\bibitem[Lin et al.(2007)]{2007Lin} Lin, D., Remillard, R.~A., \& Homan, J.\ 2007, \apj, 667, 1073, \dodoi{ doi:10.1086/521181}

\bibitem[Lin et al.(2009)]{2009Lin} Lin, D., Remillard, R.~A., \& Homan, J.\ 2009, \apj, 696, 1257,  \dodoi{doi:10.1088/0004-637X/696/2/1257}

\bibitem[Liu et al.(2001)]{2001Liu} Liu, Q.~Z., van Paradijs, J., \& van den Heuvel, E.~P.~J.\ 2001, \aap, 368, 1021, \dodoi{ doi:10.1051/0004-6361:20010075}

\bibitem[Lyubarskii(1997)]{1997Lyubarskii} Lyubarskii, Y.~E.\ 1997, \mnras, 292, 679, \dodoi{ doi:10.1093/mnras/292.3.679}

\bibitem[Maccarone et al.(2011)]{2011Maccarone} Maccarone, T.~J., Uttley, P., van der Klis, M., et al.\ 2011, \mnras, 413, 1819, \dodoi{doi:10.1111/j.1365-2966.2011.18273.x}

\bibitem[Makishima et al.(1986)]{1986Makishima} Makishima, K., Maejima, Y., Mitsuda, K., et al.\ 1986, \apj, 308, 635, \dodoi{ doi:10.1086/164534}

\bibitem[Maqbool et al.(2019)]{2019Maqbool} Maqbool, B., Mudambi, S.~P., Misra, R., et al.\ 2019, \mnras, 486, 2964, \dodoi{ doi:10.1093/mnras/stz930}

\bibitem[Matsuoka et al.(2009)]{2009Matsuoka} Matsuoka, M., Kawasaki, K., Ueno, S., et al.\ 2009, \pasj, 61, 999, \dodoi{doi:10.1093/pasj/61.5.999}

\bibitem[Misra et al.(2017)]{2017Misra} Misra, R., Yadav, J.~S., Verdhan Chauhan, J., et al.\ 2017, \apj, 835, 195, \dodoi{doi:10.3847/1538-4357/835/2/195}

\bibitem[Misra(2000)]{2000Misra} Misra, R.\ 2000, \apjl, 529, L95, \dodoi{doi:10.1086/312470}

\bibitem[Mitsuda et al.(1984)]{1984Mitsuda} Mitsuda, K., Inoue, H., Koyama, K., et al.\ 1984, \pasj, 36, 741

\bibitem[Mitsuda et al.(1989)]{1989Mitsuda} Mitsuda, K., Inoue, H., Nakamura, N., et al.\ 1989, \pasj, 41, 97

\bibitem[Mondal et al.(2019)]{2019Mondal} Mondal, A.~S., Dewangan, G.~C., \& Raychaudhuri, B.\ 2019, \mnras, 487, 5441, \dodoi{ doi:10.1093/mnras/stz1658}

\bibitem[Motta(2016)]{2016Motta} Motta, S.~E.\ 2016, Astronomische Nachrichten, 337, 398, \dodoi{ doi:10.1002/asna.201612320}

\bibitem[Mudambi et al.(2020)]{2020Mudambi} Mudambi, S.~P., Maqbool, B., Misra, R., et al.\ 2020, \apjl, 889, L17, \dodoi{ doi:10.3847/2041-8213/ab66bc}

\bibitem[Murakami et al.(1980)]{1980Murakami} Murakami, T., Inoue, H., Koyama, K., et al.\ 1980, \apjl, 240, L143, \dodoi{doi:10.1086/183341}

\bibitem[Nath et al.(2022)]{2022Nath} Nath, A., Sarkar, B., Roy, J., et al.\ 2022, Journal of Astrophysics and Astronomy, 43, 93, \dodoi{ doi:10.1007/s12036-022-09878-9}

\bibitem[Nowak(1995)]{1995Nowak} Nowak, M.~A.\ 1995, \pasp, 107, 1207, \dodoi{doi:10.1086/133679}


\bibitem[{\"O}zel et al.(2016)]{2016Ozel} {\"O}zel, F., Psaltis, D., G{\"u}ver, T., et al.\ 2016, \apj, 820, 28,\dodoi{ doi:10.3847/0004-637X/820/1/28}

\bibitem[Paizis et al.(2006)]{2006Paizis} Paizis, A., Farinelli, R., Titarchuk, L., et al.\ 2006, \aap, 459, 187, \dodoi{ doi:10.1051/0004-6361:20065792}

\bibitem[Penninx et al.(1989)]{1989Penninx} Penninx, W., Damen, E., Tan, J., et al.\ 1989, \aap, 208, 146

\bibitem[Popham \& Sunyaev(2001)]{2001Popham} Popham, R. \& Sunyaev, R.\ 2001, X-ray Astronomy: Stellar Endpoints, AGN, and the Diffuse X-ray Background, 599, 870, \dodoi{ doi:10.1063/1.1434763}

\bibitem[Poutanen et al.(2014)]{2014Poutanen} Poutanen, J., N{\"a}ttil{\"a}, J., Kajava, J.~J.~E., et al.\ 2014, \mnras, 442, 3777, \dodoi{ doi:10.1093/mnras/stu1139}

\bibitem[Prins \& van der Klis(1997)]{1997Prins} Prins, S. \& van der Klis, M.\ 1997, \aap, 319, 498, \dodoi{doi:10.48550/arXiv.astro-ph/9701149}

\bibitem[Psaltis et al.(1999)]{1999Psaltis} Psaltis, D., Belloni, T., \& van der Klis, M.\ 1999, \apj, 520, 262, \dodoi{doi:10.1086/307436}

\bibitem[Remillard et al.(2022)]{2022Remillard} Remillard, R.~A., Loewenstein, M., Steiner, J.~F., et al.\ 2022, \aj, 163, 130, \dodoi{ doi:10.3847/1538-3881/ac4ae6}

\bibitem[Roy et al.(2021)]{2021Roy} Roy, P., Beri, A., \& Bhattacharyya, S.\ 2021, \mnras, 508, 2123, \dodoi{doi:10.1093/mnras/stab2680}

\bibitem[Roy et al.(2022)]{2022Roy} Roy, P., Beri, A., \& Mondal, A.~S.\ 2022, Journal of Astrophysics and Astronomy, 43, 45, \dodoi{ doi:10.1007/s12036-022-09825-8}

\bibitem[Shaposhnikov et al.(2003)]{2003Shaposhnikov} Shaposhnikov, N., Titarchuk, L., \& Haberl, F.\ 2003, \apjl, 593, L35, \dodoi{ doi:10.1086/378255}

\bibitem[Shimura \& Takahara(1995)]{1995Shimura} Shimura, T. \& Takahara, F.\ 1995, \apj, 445, 780, \dodoi{ doi:10.1086/175740}

\bibitem[Singh et al.(2014)]{2014Singh} Singh, K.~P., Tandon, S.~N., Agrawal, P.~C., et al.\ 2014, \procspie, 9144, 91441S, \dodoi{ doi:10.1117/12.2062667}

\bibitem[Singh et al.(2016)]{2016Singh} Singh, K.~P., Stewart, G.~C., Chandra, S., et al.\ 2016, \procspie, 9905, 99051E, \dodoi{ doi:10.1117/12.2235309}

\bibitem[Singh et al.(2017)]{2017Singh} Singh, K.~P., Stewart, G.~C., Westergaard, N.~J., et al.\ 2017, Journal of Astrophysics and Astronomy, 38, 29 \dodoi{doi:10.1007/s12036-017-9448-7}

\bibitem[Tananbaum et al.(1976)]{1976Tananbaum} Tananbaum, H., Chaisson, L.~J., Forman, W., et al.\ 1976, \apjl, 209, L125, \dodoi{ doi:10.1086/182282}

\bibitem[Turner \& Breedon(1984)]{1984Turner} Turner, M.~J.~L. \& Breedon, L.~M.\ 1984, \mnras, 208, 29P, \dodoi{ doi:10.1093/mnras/208.1.29P}

\bibitem[van der Klis et al.(1990)]{1990vander} van der Klis, M., Hasinger, G., Damen, E., et al.\ 1990, \apjl, 360, L19, \dodoi{doi:10.1086/185802}
\bibitem[van der Klis(1989a)]{1989vanderklis} van der Klis, M.\ 1989a, Timing Neutron Stars, 262, 27, \dodoi{doi:10.1007/978-94-009-2273-0\_3}

\bibitem[van der Klis(1989b)]{1989van} van der Klis, M.\ 1989b, \araa, 27, 517, \dodoi{doi:10.1146/annurev.aa.27.090189.002505}

\bibitem[van der Klis(1989c)]{1989vander} van der Klis, M.\ 1989c, Two Topics in X-Ray Astronomy, 1, 203.

\bibitem[van der Klis(1994)]{1994vander} van der Klis, M.\ 1994, \aap, 283, 469

\bibitem[van der Klis(1994b)]{1994Van} van der Klis, M.\ 1994, \apjs, 92, 511. doi:10.1086/192006

\bibitem[van der Klis(1994)]{1995vander} van der Klis, M. 1995, in {\it X-Ray Binaries}, Lewin W. H. G., van Paradijs 
J., van den Heuvel E. P. J. eds., Cambridge Astrophysics Series, Cambridge, p. 252

\bibitem[van der Klis(2000)]{2000vander} van der Klis, M.\ 2000, \araa, 38, 717, \dodoi{doi:10.1146/annurev.astro.38.1.717}

\bibitem[van der Klis(2004)]{2004Vander} van der Klis, M.\ 2004, astro-ph/0410551, \dodoi{doi:10.48550/arXiv.astro-ph/0410551}

\bibitem[van der Klis(2006a)]{2006van} van der Klis, M.\ 2006a, Advances in Space Research, 38, 2675, \dodoi{doi:10.1016/j.asr.2005.11.026}

\bibitem[van der Klis(2006b)]{2006vander} van der Klis, M.\ 2006b, Compact stellar X-ray sources (Cambridge: Cambridge Univ. Press)

\bibitem[Vaughan \& Nowak(1997)]{1997Vaughan} Vaughan, B.~A. \& Nowak, M.~A.\ 1997, \apjl, 474, L43. doi:10.1086/310430

\bibitem[van Straaten et al.(2003)]{2003Straaten} van Straaten, S., van der Klis, M., \& M{\'e}ndez, M.\ 2003, New Views on Microquasars, 52, \dodoi{ doi:10.48550/arXiv.astro-ph/0207646}

\bibitem[White et al.(1988)]{1988White} White, N.~E., Stella, L., \& Parmar, A.~N.\ 1988, \apj, 324, 363, \dodoi{ doi:10.1086/165901}

\bibitem[Wijnands et al.(1997)]{1997Wijnands} Wijnands, R., Homan, J., van der Klis, M., et al.\ 1997, \apjl, 490, L157, \dodoi{ doi:10.1086/311039}

\bibitem[Wijnands \& van der Klis(1999)]{1999Wijnands} Wijnands, R. \& van der Klis, M.\ 1999, \apj, 514, 939, \dodoi{ doi:10.1086/306993}

\bibitem[Wijnands et al.(1999)]{1999Wij} Wijnands, R., Homan, J., \& van der Klis, M.\ 1999, \apjl, 526, L33, \dodoi{doi:10.1086/312365}


\bibitem[Wilms et al.(2000)]{2000Wilms} Wilms, J., Allen, A., \& McCray, R.\ 2000, \apj, 542, 914, \dodoi{doi:10.1086/317016}

\bibitem[Yadav et al.(2016)]{2016Yadav} Yadav, J.~S., Misra, R., Verdhan Chauhan, J., et al.\ 2016, \apj, 833, 27, \dodoi{doi:10.3847/0004-637X/833/1/27}

\bibitem[Yang et al.(2022)]{2022Yang} Yang, Z.-X., Zhang, L., Bu, Q.-C., et al.\ 2022, \apj, 932, 7, \dodoi{doi:10.3847/1538-4357/ac63af}

\bibitem[Yoshida et al.(1993)]{1993Yoshida} Yoshida, K., Mitsuda, K., Ebisawa, K., et al.\ 1993, \pasj, 45, 605

\bibitem[Yu et al.(2003)]{2003Yu} Yu, W., van der Klis, M., \& Fender, R.~P.\ 2003, New Views on Microquasars, 80, \dodoi{doi:10.48550/arXiv.astro-ph/0207645}

\bibitem[Zdziarski et al.(1996)]{1996Zdziarski} Zdziarski, A.~A., Johnson, W.~N., \& Magdziarz, P.\ 1996, \mnras, 283, 193, \dodoi{ doi:10.1093/mnras/283.1.193}

\bibitem[Zdziarski et al.(2020)]{2020Zdziarski} Zdziarski, A.~A., Szanecki, M., Poutanen, J., et al.\ 2020, \mnras, 492, 5234, \dodoi{ doi:10.1093/mnras/staa159}

\bibitem[Zhang et al.(1995)]{1995Zhang} Zhang, W., Jahoda, K., Swank, J.~H., et al.\ 1995, \apj, 449, 930, \dodoi{doi:10.1086/176111}

\bibitem[Zhang et al.(1996)]{1996Zhang} Zhang, S.~N., Harmon, B.~A., Paciesas, W.~S., et al.\ 1996, \aaps, 120, 279

\bibitem[{\.Z}ycki et al.(1999)]{1999Zycki} {\.Z}ycki, P.~T., Done, C., \& Smith, D.~A.\ 1999, \mnras, 309, 561, \dodoi{ doi:10.1046/j.1365-8711.1999.02885.x}

\end{thebibliography}
\bibliographystyle{aasjournal}

\end{document}